\documentclass[preprint2]{aastex62}
\usepackage{bm}
\usepackage{mathrsfs}
\usepackage{amssymb}
\usepackage{amsmath}
\usepackage{graphicx}
\usepackage{array}
\usepackage{color}
\usepackage{tabularx}
\usepackage{lineno}
\usepackage{float}

\newcommand{\tttt}[1]{{$\times 10^{#1}$}} 
\newcommand{\fgw}[1]{{$f_{\mathrm{GW}}$\,}}

\begin{document}
\title{Modeling the Galactic Neutron Star Population for Use in Continuous Gravitational Wave Searches}


\author[0000-0002-7775-5423]{Brendan T. Reed}\email{reedbr@iu.edu}
\affiliation{Department of Astronomy, Indiana University,
                  Bloomington, IN 47405, USA}
\affiliation{Center for Exploration of Energy and Matter and
                  Department of Physics, Indiana University,
                  Bloomington, IN 47405, USA}

\author[0000-0002-2531-8854]{Alex Deibel}\email{adeibel@iu.edu}
\affiliation{Department of Astronomy, Indiana University,
                  Bloomington, IN 47405, USA}

\author[0000-0001-7271-9098]{C. J. Horowitz}\email{horowit@indiana.edu}
\affiliation{Center for Exploration of Energy and Matter and
                  Department of Physics, Indiana University,
                  Bloomington, IN 47405, USA}

\date{\today}
\begin{abstract}
Searches for continuous gravitational waves from \textit{unknown} Galactic neutron stars provide limits on the shapes of neutron stars. A rotating neutron star will produce gravitational waves if asymmetric deformations exist in its structure that are characterized by the star's ellipticity. In this study, we use a simple model of the spatial and spin distribution of Galactic neutron stars to estimate the total number of neutron stars probed, using gravitational waves, to a given upper limit on the ellipticity. This may help optimize future searches with improved sensitivity.  The improved sensitivity of third-generation gravitational wave detectors may increase the number of neutron stars probed, to a given ellipticity, by factors of 100 to 1000.
\end{abstract}

\section{Introduction}\label{sec:intro}
Within an order of magnitude, the age of the Milky Way is $\sim 10^{10}\, \mathrm{yrs}$ and has a Galactic supernovae rate of $\sim 1$ per century \citep{diehl:2006}. We can therefore estimate that $N_0\sim 10^8$ neutron stars (NS) have been born in our galaxy to-date. Of that number, a relatively small fraction are known through electromagnetic searches \textrm{--} a few thousand mostly radio pulsars \textrm{--} see, for example, the ATNF Pulsar Database \citep{ATNF,ATNF-site}. Gravitational waves (GW) may be a means to discover some of the remaining unknown NSs and study the distribution of their shapes.



Any rotating NS with asymmetric deformations will produce continuous gravitational waves (CGWs) via quadrupole radiation \citep{zimmerman:1979,lasky:2015} and the observed background of CGWs from GW detectors may reveal unknown NSs \citep{PhysRevD.100.064013,all-sky:2019}.  A rotating NS radiates CGWs with strain amplitude $h_0$ according to
\begin{eqnarray}
h_0 = \frac{4\pi^2G}{c^4}\frac{I_{zz}f_{\rm GW}^2}{d}\epsilon \ ,
\label{epsilon}
\end{eqnarray} where $d$ is the distance to the source and the gravitational wave frequency is $f_{\rm GW} = 2 \nu$ for a NS rotating with spin frequency $\nu$ \citep{riles:2017,all-sky:2019}. This relation is notable in that it is linearly dependent on the NS's ellipticity, 
\begin{eqnarray}
\epsilon=\sqrt{\frac{8\pi}{15}}\frac{Q_{22}}{I_{zz}}=\frac{I_{xx}-I_{yy}}{I_{zz}},
\label{eq:ell}
\end{eqnarray} defined here in either terms of the quadrupole moment $Q_{22}$ or fractional difference in principle moments of inertia \citep{owen:2005,riles:2017}. 

We expect a distribution of ellipticities across Galactic NSs. The maximum allowed ellipticity may be limited by the breaking strain of the NS crust to $\epsilon\lesssim {\rm few}\times 10^{-6}$ \citep{ushomirsky:2000,PhysRevLett.102.191102,10.1093/mnras/staa3635}. In order for a NS to support a larger $\epsilon$, there may need to be an exotic solid phase in the core, such as crystalline quark matter \citep{owen:2005,johnson-mcdaniel:2013}. Constraining the ellipticity further, observations of millisecond pulsars (MSPs) suggest that the NS ellipticity reaches a minimum near $\epsilon \approx 10^{-9}$ \citep{woan:2018}.

CGWs from Galactic NS are expected to be $\sim 10^{-4}$ times lower in amplitude than the GW signal from the binary mergers of compact objects \citep{riles:2017}. There have been many CGW searches from known pulsars, see for example \citep{PhysRevD.99.122002}.  Furthermore, one can gain sensitivity to weak CGW signals by integrating for a long time. For \textit{known} pulsars, however, radio or X-ray spin-down luminosity place an observational limit on the power in GW radiation. Alternatively, there are a number of all sky searches for CGWs from {\it unknown} NSs \citep{all-sky:2019,steltner2020einsteinhome,dergachev:2020,Dergachev_Papa:2021a,Dergachev_Papa:2021b}. An unknown NS could be a strong CGW source with an unconstrained spin-down power. However, to-date no CGW signals have been detected. 

The physics implications of negative CGW searches are presently unclear, but we can use these result to infer some interesting limits on the NS population. Because the CGW strain amplitude depends inversely on the distance to the NS ($h_0 \sim 1/d$), the lack of detected CGWs constrains the NS population within that distance from Earth. Of course, the strain amplitude also depends on the spin-frequency of the NS producing the GWs as $f_{\mathrm GW} = 2 \nu$ and the ellipticity of the NS ($h_0 \sim f^2_{\mathrm GW} \epsilon$). Assuming a spatial distribution and spin distribution for Galactic NSs near Earth, we can therefore infer limits on the ellipticities of the NS population producing CGWs within a distance $d$ from Earth. Doing so would allow us to estimate \emph{how many NSs are actually being probed by a given CGW search.}

In this paper, we develop a very simple population model of the spatial and spin distribution of Galactic NSs. We then use this model to estimate the number of Galactic NSs probed by recent CGW searches. We detail our distribution choices in Sec.~\ref{sec:modelling}. These distributions can be used to help optimize future searches and to infer the physics implications of search results. In Sec.~\ref{sec:results} we use our NS distribution, along with search limits on $h_0$, to infer a distribution of upper limits on NS ellipticities. Lastly, we discuss the implications and conclusions of our findings and possible future studies to improve these limits in Sec.~\ref{sec:discussion}. We also discuss our assumptions and their impact on the results; for example, the time-dependence of the CGW source, the effects of binary pairs of NS, and the NS birth history in the Milky Way. 

\section{Modeling Neutron Star Distributions}
\label{sec:modelling}
In this section, we develop a simple model for the distribution of NSs in the galaxy to provide a simple first estimate of the number of NSs probed to a variety of ellipticities via CGW data. We explain our calculations of the maximum distance from earth that is probed at a given frequency in \autoref{sec:strain}. We then give our assumptions and calculations for the spatial distribution of NSs in \autoref{sec:dist} and detail our choice of spin-frequency distribution in \autoref{sec:spin}. Finally, we show the calculation of the unknown NS population in \autoref{sec:popest}.
\subsection{Gravitational Wave Strain Data}
\label{sec:strain}
\autoref{eq:ell} gives the definition of $\epsilon$, characterized by an asymmetric deformation on the surface of a NS. This asymmetry will cause a \textit{rapidly} rotating NS to emit CGWs with a strain amplitude given by \autoref{epsilon}. The strain amplitude $h_0$ is sensitive to the frequency of the GW signal and has a complicated behavior. 

\begin{figure*}[t]
    \includegraphics[width=\textwidth]{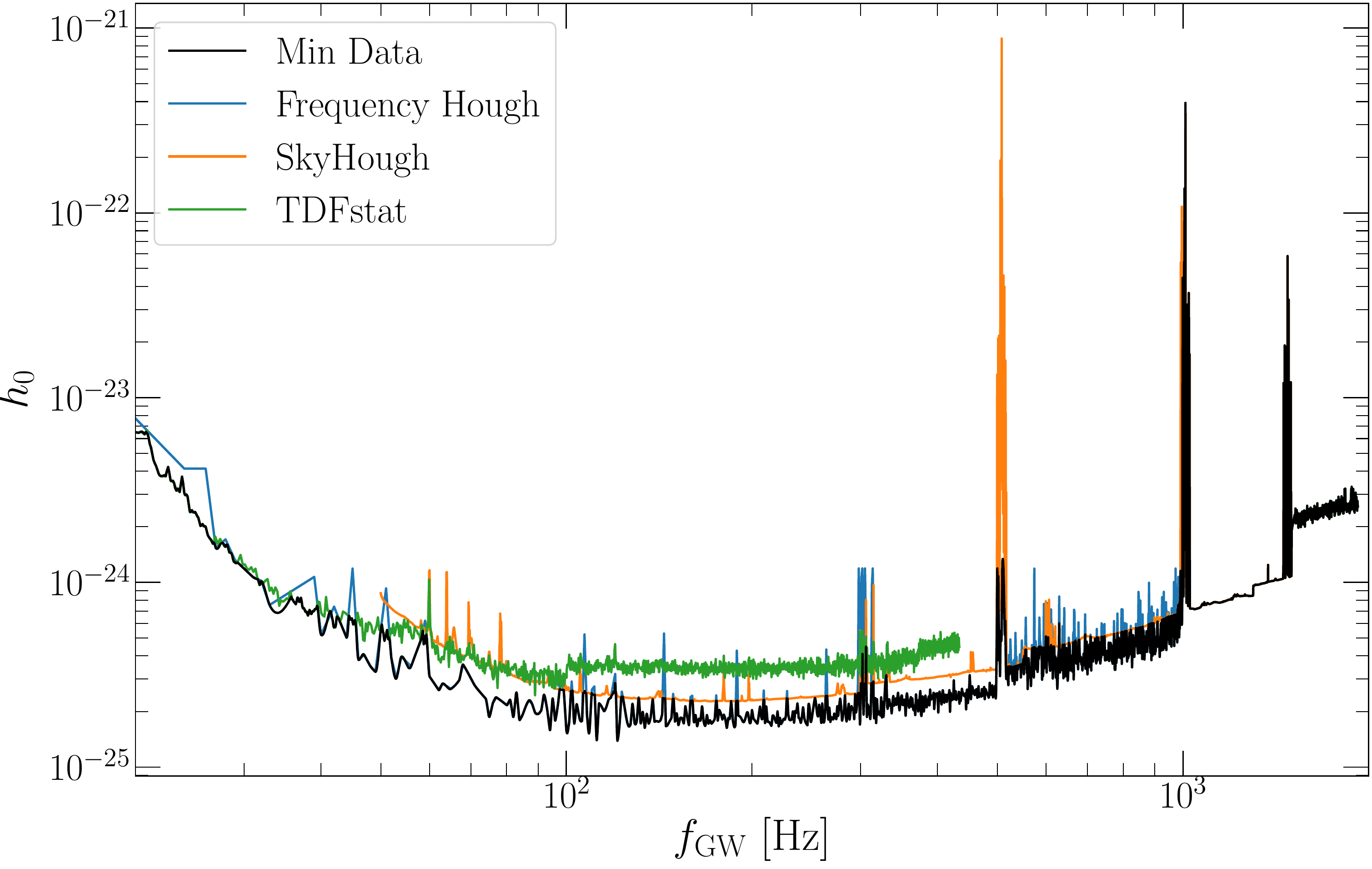}
    \caption{Strain amplitude versus frequency used in this work. The three colored lines represent the 95\% confidence level obtained for each of the three pipelines used in \cite{all-sky:2019}. The black line is the data that we use in the calculation of $d(f_{\mathrm GW},\epsilon)$, obtained by taking the smallest value for $h_0$ at each frequency from among the three pipelines.}
    \label{fig:strain_data}
\end{figure*}

We use data from \citet{all-sky:2019} which presented a detailed analysis of their CGW search limits on $h_0$ as a function of $f_{\rm GW}$ at 95\% confidence. Within this data are the constraints from the three pipelines \textit{SkyHough} \citep{Covas:2004}, \textit{Frequency-Hough} \citep{FH}, and \textit{TDFstat} \citep{TDFstat} which have different sensitivities in the range of frequencies considered by \cite{all-sky:2019}. Because there exists some overlap in the strain among the three pipelines, we define a grid of 20-1922 Hz and take the smallest of the three's $h_0$ at each frequency to use in our calculations. We plot this and the data from the three pipelines in \autoref{fig:strain_data}.

\begin{figure}[htb]
    \centering
    \includegraphics[width=\columnwidth]{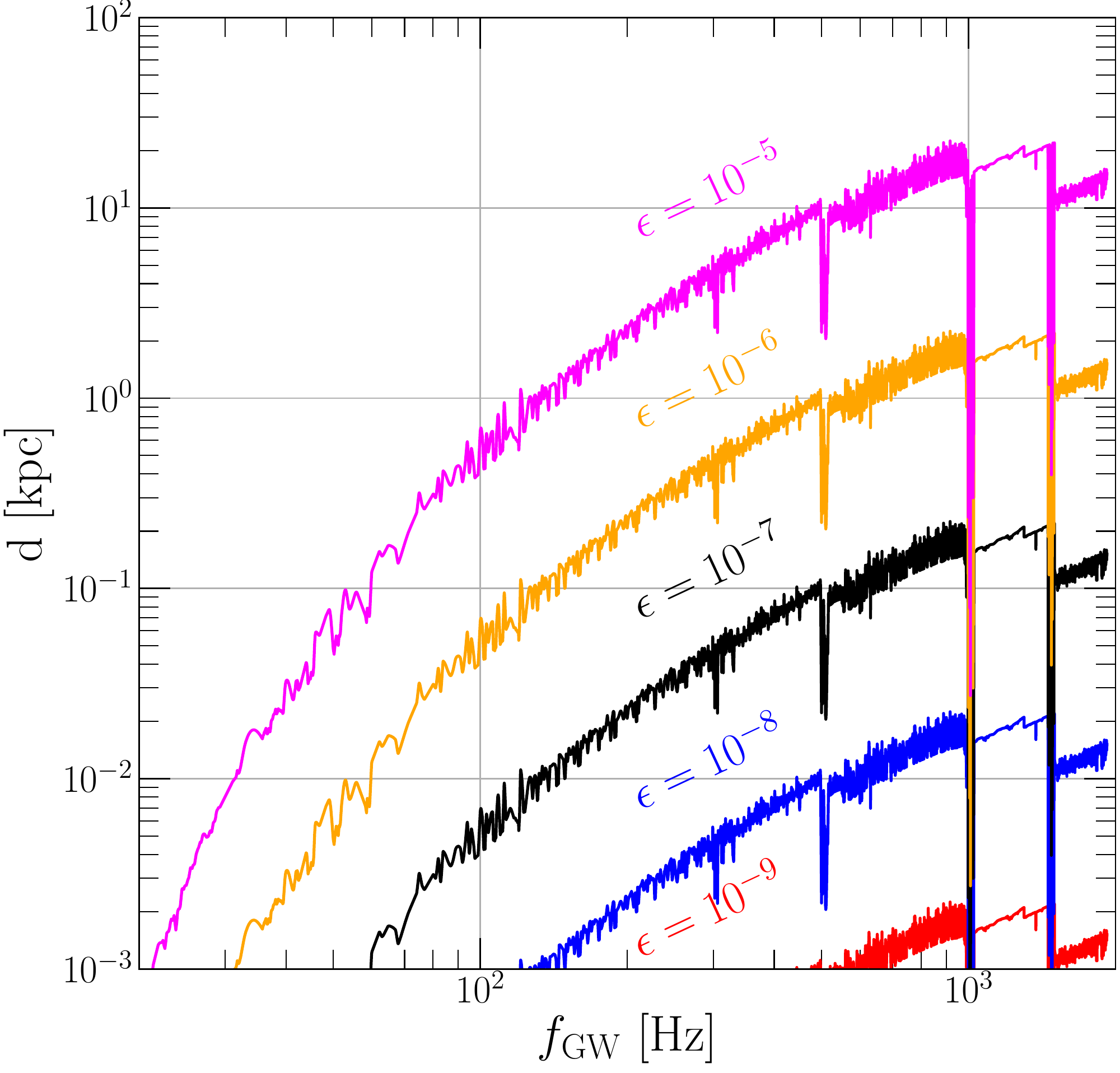}
    \caption{Maximum distance from Earth probed at a given GW frequency for a range of possible NS ellipticities. The curves shown here are the result of using $h_0(f_{\rm GW})$ tabulated in \autoref{fig:strain_data} which is then substituted into \autoref{eq:dist} for different $\epsilon$.}
    \label{fig:d-f}
\end{figure}

By simple inversion of \autoref{epsilon}, we can solve for the maximum distance from Earth to which a NS with frequency $f_{\rm GW}$ and ellipticity $\epsilon$ has been excluded. Explicitly,
\begin{eqnarray}
d(f_{\rm GW},\epsilon) = \frac{4\pi^2 G}{c^4} \frac{I_{zz}f^2_{\rm GW}\epsilon}{h_0(f_{\rm GW})} \, .
\label{eq:dist}
\end{eqnarray}
The distance can be easily obtained by fixing a desired value for $\epsilon$, using a canonical value for $I_{zz}=10^{45}\,\mathrm{g \, cm^2}$, and then choosing a desired $f_{\mathrm{GW}}$ value. For illustration, we plot the distance versus $f_{\rm GW}$ for various $\epsilon$ values in \autoref{fig:d-f}. Note that current GW interferometers are insensitive below 20 Hz.

Because the maximum distance in \autoref{fig:d-f} goes as $d \sim f_{\mathrm{GW}}^2 \epsilon$, the greatest distances probed are for $f_{\mathrm{GW}} \gtrsim 1,000 \, \mathrm{Hz}$ and large ellipticity $\epsilon \sim 10^{-5}$. In particular, NSs with $\epsilon \sim 10^{-5}$ are excluded to around $d\approx 20\, \mathrm{kpc}$ (likely the entire Galactic disk) for $f_{\mathrm{GW}} \gtrsim 1,000 \, \mathrm{Hz}$ and to $d\approx 500 \, \mathrm{pc}$ for $f_{\mathrm{GW}} = 100 \, \mathrm{Hz}$. For NSs with $\epsilon \sim 10^{-6}$, closer to the breaking strain of the crust, they are excluded to approximately $d\approx 2 \, \mathrm{kpc}$ for $f_{\mathrm{GW}} \gtrsim 1,000 \, \mathrm{Hz}$ and around $d\approx 50 \, \mathrm{pc}$ for $f_{\mathrm{GW}} = 100 \, \mathrm{Hz}$. For NSs with $\epsilon \sim 10^{-9}$ they are excluded up to $d \approx 2 \, \mathrm{pc}$ for $f_{\mathrm{GW}} \gtrsim 1,000 \, \mathrm{Hz}$ and $d << 1 \, \mathrm{pc}$ for $f_{\mathrm{GW}} = 100 \, \mathrm{Hz}$.

\subsection{Spatial Distribution}
\label{sec:dist}
The NSs that fall within the range of the GW detector as defined by \autoref{eq:dist} reside in the Galactic disk. The spatial distribution of Galactic NSs is believed to approximately follow an exponential distribution in the vertical direction above the disk and a Gaussian-like distribution in the radial direction \citep{Binney:1998,faucher:2010,Taani:2012}. We shall adopt the following equation for the 3D-density of neutron stars in the Galaxy
\begin{eqnarray}
\rho_c(r_c,z) = \frac{N_0}{4\pi\sigma_r^2 z_0}\exp\left[-\frac{r_c^2}{2\sigma_r^2}\right]\exp\left[-\frac{|z|}{z_0}\right],
\label{3d_dist}
\end{eqnarray}
where $r_c$ is the cylindrical radius from the Galactic Center, $\sigma_r$ is a radius parameter, $N_0$ is the total number of NSs, and $z_0$ is the disk thickness. For $\sigma_r$ we adopt a value of 5 kpc as in \cite{faucher:2010} and we use $N_0=10^8$ as discussed in \autoref{sec:intro}. 
For normal stars, $z_0$ is often in the range of 0.5-1.0 kpc \citep{Binney:1998,Dynamics}. However, NSs may follow a different distribution due to supernova kicks. Therefore, to probe a wider range of models for the distribution, we choose to vary $z_0$ for the values in \autoref{tab:constants}.

\begin{table}[htb]
    \centering
    \begin{tabular}{c|c|c}\hline\hline
        Parameter  & Symbol & Adopted Value(s)\\\hline
        Radius Parameter & $\sigma_r$ & 5 kpc \\
        Disk Thickness & $z_0$ & 0.1, 2.0, 4.0 kpc \\
        Distance to GC & $R_e$ & 8.25 kpc \\
        Normalization & $N_0$ & $10^8$ stars\\\hline
    \end{tabular}
    \caption{Adopted parameter values for the population distributions used in this study.}
    \label{tab:constants}
\end{table}

\begin{figure*}[htb!]
    \centering
    \includegraphics[width=\textwidth]{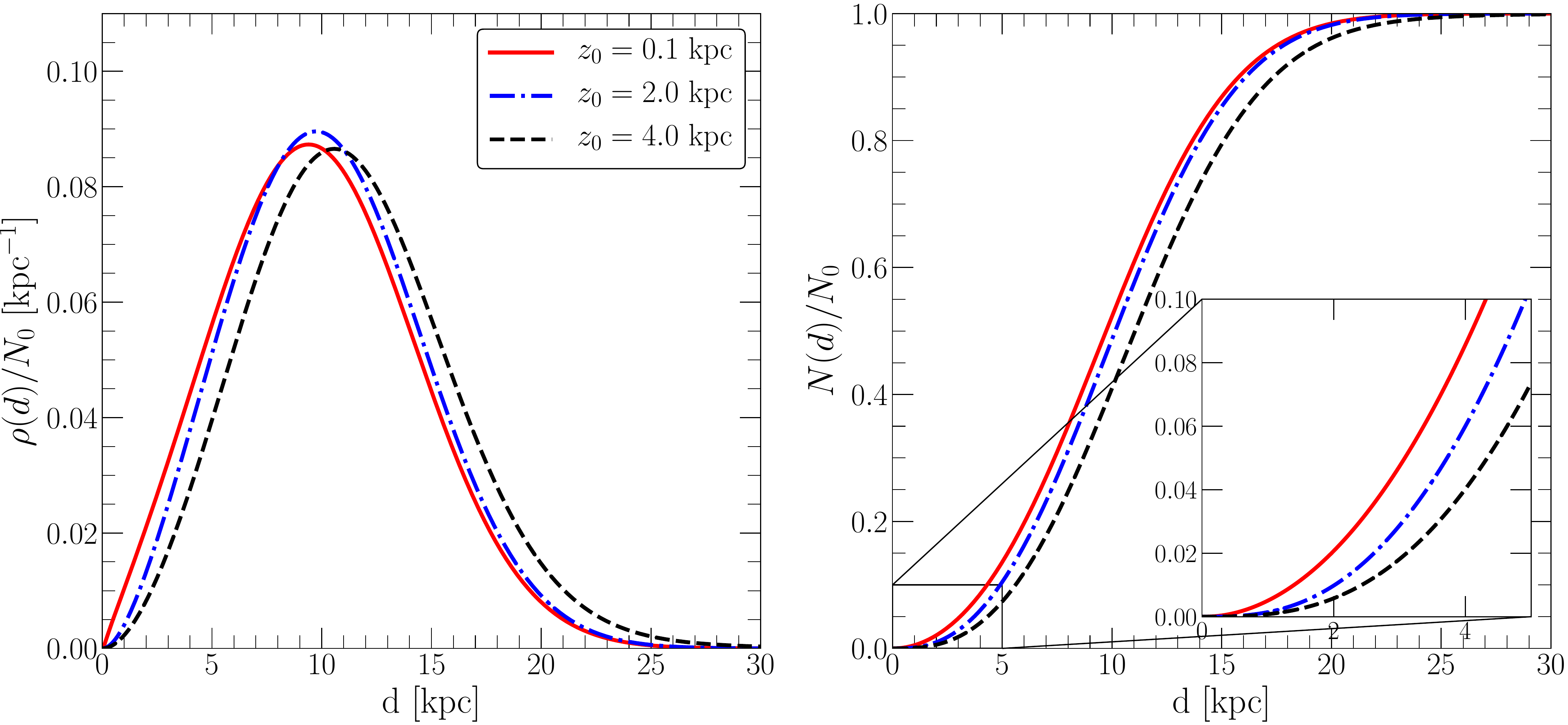}
    \caption{\textit{Left}: 1D probability density distribution at three different values for the scale height, see \autoref{cdf}. \textit{Right}: Cumulative distribution function of NS, calculated by integrating the 1D density from 0 to $d$. Inset shows the spread in models' behavior at low values of $d$.}
    \label{num_dens}
\end{figure*}

We then perform a coordinate transformation from the cylindrical $(r_c,z)$ to the average 3D distance from Earth, $d$. This can be done by first transforming $r_c$ to be centered on the Earth via $\Vec{r_c}\longrightarrow \Vec{r}+\Vec{R_e}$ where $R_e = 8.25 \, \textrm{kpc}$ is the distance from the Galactic Center to Earth \citep{galcenter-dist}.  Plugging this substitution into \autoref{3d_dist} and integrating over the angular direction gives us

\begin{equation}
\rho'(r,z) = \frac{N_0{\rm e}^{-\frac{|z|}{z_0}}}{2\sigma_r^2z_0}I_0\left(\frac{R_e r}{\sigma_r^2}\right)\exp\left[-\frac{(r^2+R_e^2)}{2\sigma_r^2}\right]
\end{equation}
where $I_0$ is the modified Bessel function. We note that this distribution is normalized to $N_0$,
\begin{eqnarray}
\int_{-\infty}^\infty dz \int_0^\infty rdr\rho'(r,z)= N_0
\end{eqnarray}

Now, using $d$ for the 3D distance from Earth, $d = \sqrt{r^2+z^2}$, we can then arrive at an average 1D density $\rho(d)$ by performing the following integral
\begin{equation}
\rho(d)=\int_{-\infty}^\infty dz\int_0^\infty rdr  \rho'(r,z)\delta(\sqrt{r^2+z^2}-d)\, .
\end{equation}
Integrating over the radial coordinate first, we then recast $z$ in terms of a scaled variable $x=z/d$. With this, we have arrived at the probability density distribution
\begin{align}
\rho(d) =& \frac{N_0d^2}{\sigma_r^2 z_0}\int_0^1  \exp\left[-\frac{xd}{z_0}\right]
I_0\left[\frac{R_ed\sqrt{1-x^2}}{\sigma_r^2}\right]\nonumber\\& \times\exp\left[-\frac{R_e^2+d^2(1-x^2)}{2\sigma_r^2}\right]dx\, ,
\label{cdf}
\end{align}
which gives the likelihood that a NS is a distance $d$ from Earth. We plot the probability density distribution within $30 \, \mathrm{kpc}$ of Earth in the left panel of \autoref{num_dens}. Finally, we integrate \autoref{cdf} to arrive at the cumulative distribution function $N(d)$ at a distance $d$, defined here as 
\begin{eqnarray}
N(d) = \int_0^d \rho(y) dy
\end{eqnarray}
We plot $N(d)$ in the right panel of \autoref{num_dens}.

\subsection{Spin-frequency Distribution}
\label{sec:spin}

A neutron star spinning with frequency $\nu$ emits CGWs with frequency $f_{\rm GW} = 2\nu$ according to \autoref{epsilon}. The observed frequency of many MSPs is believed to be due to spin-up in a NS's low-mass X-ray binary phase \citep[e.g. ][]{Radhakrishnan:1982,Wijnands:1998,Papitto:2013}. Over the lifetime of this phase, asymmetric electron-capture reaction layers on the accreting neutron star may lead to an asymmetric deformation \citep{bildsten:1998,ushomirsky:2000} because of an asymmetry in the temperature distribution of the NS's magnetic field. The observed distribution of $\nu$ largely depends on the spin evolution in this phase \citep{Bhattacharyya:2021}. After this phase concludes, the spinning NS continues to emit CGWs which affects both the time evolution of $\epsilon$ and $\nu$.

\begin{figure*}[htb]
    \includegraphics[width=\textwidth]{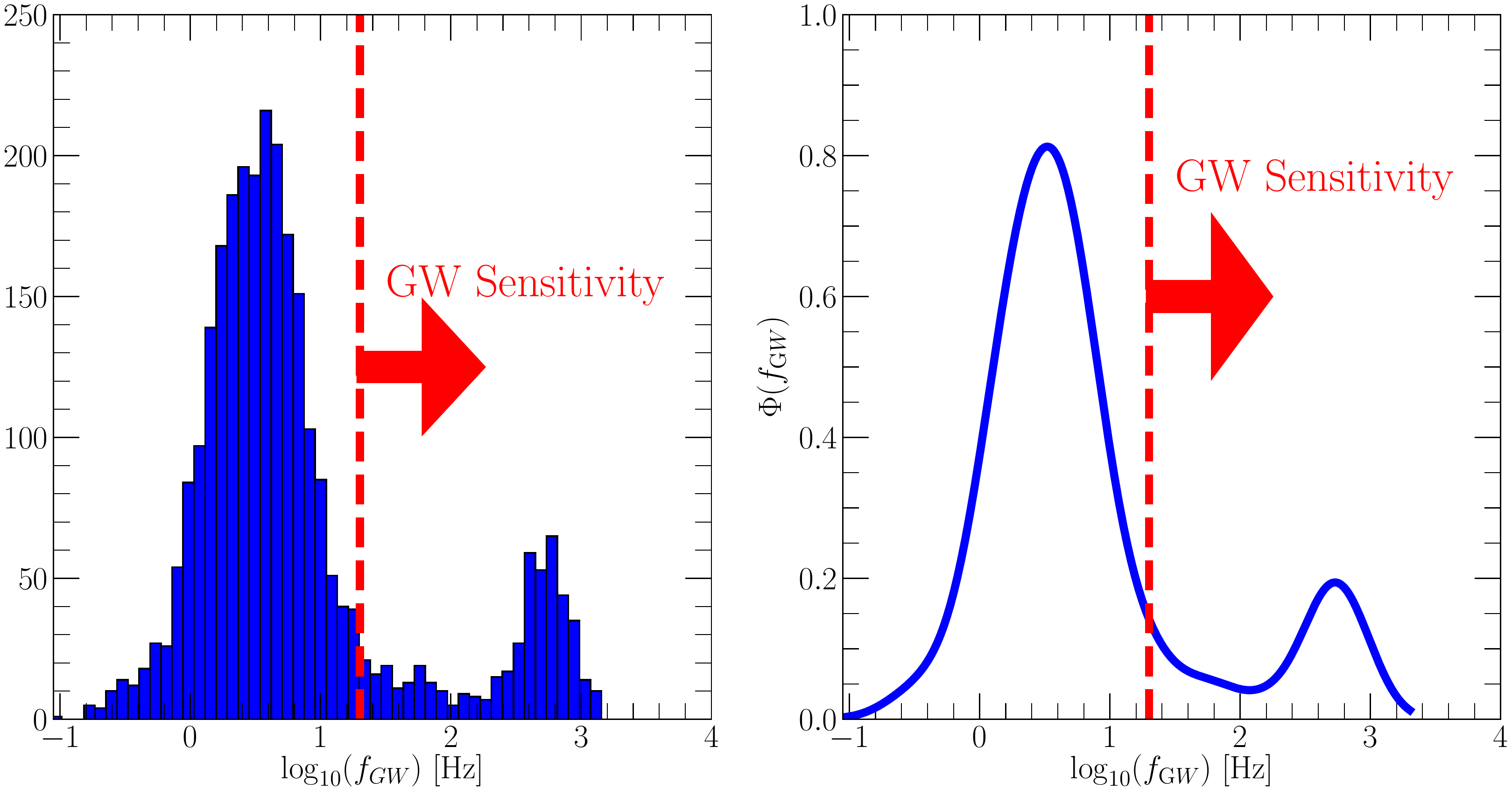}
    \caption{
    \textit{Left}: Histogram of the distribution of $f_{\rm GW}$ from pulsars in the ATNF pulsar database. Here, we convert the normal spin-frequency $\nu$ of the NS's rotation to the GW frequency via $f_{\rm GW}=2\nu$. \textit{Right}: Calculated probability density function of the histogram on the left, normalized to unity. The 20 Hz limit of GW detector sensitivity is shown on both by the vertical red line and arrow to indicate direction of limit.
    }
    \label{fig:kde}
\end{figure*}

As a simple starting point, we assume that the true distribution of Galactic NS spin-frequencies is the same as the observed spin-frequency distribution of 2811 pulsars from the ATNF Pulsar Database \citep{ATNF,ATNF-site}. In the left panel of \autoref{fig:kde}, we show a histogram of the $f_{GW}$ expected from pulsars in the database. Within the sample, there are 489 pulsars with a spin frequency above $10$ Hz that produce $f_{\mathrm{GW}} > 20 \, \mathrm{Hz}$ and fall within GW detector sensitivity. We note the maximally rotating pulsar in the catalog is PSR J1748-2446ad \citep{Hessels:2006} and has $\nu \approx 716$ Hz ($f_{\mathrm{GW}} \approx 1432 \, \mathrm{Hz}$), which is the fastest rotating pulsar yet observed.

Using a Kernel-Density Estimator (KDE) from \citet{scipy} (package documentation in \cite{scipy.kde}) we calculate a probability distribution function (PDF) of $f_{\rm GW}$ from the pulsar distribution, $\Phi$. This method uses a Gaussian with Scott's rule \citep{scottsrule} of $n^{-1/5}$ and is convenient as it produces a continuous function of $f_{\mathrm GW}$ which makes solving integrals with it much easier. We then normalize $\Phi$ to unity
\begin{eqnarray}
1 = \int_{-\infty}^\infty \Phi(\log f)d(\log f)
\end{eqnarray}
giving us a normalized PDF of \fgw. We note that $\Phi(f)$ is continuous, but drops off quickly after $f_{\mathrm GW}\gtrsim1430$ Hz because there are no observed pulsars with spin frequencies above $\nu = 716 \, \mathrm{Hz}$. Above 2000 Hz, we set $\Phi(f>2000~\rm{Hz})=0$ since the data from \autoref{sec:strain} does not go above $f_{\mathrm GW}=2000$ Hz. We show the PDF in the right panel of \autoref{fig:kde}. Note that the KDE was fit in $\log_{10}(f_{\rm GW})$ to achieve higher accuracy at low values of $f_{\rm GW}$. Henceforth, all $\log_{10}$ shall be shortened to just $\log$.

\subsection{Unknown NS Population Estimate}
\label{sec:popest}

Ground-based GW detectors are insensitive for $f_{\rm GW}<20$ Hz and based on our catalog, $\gtrsim 83\%$ of \textit{known} pulsars are therefore spinning too slowly to be detectable via CGWs. As a first-order estimate we can say that CGW searches are at most sensitive to the remaining $17\%$, that is, approximately $\sim17$ million NSs throughout the Galaxy,
\begin{eqnarray}
0.17\approx \int_{\mathrm \log(20 Hz)}^{\infty} \Phi(\log f)d(\log f).
\end{eqnarray}
In practice, ground-based GW detector sensitivity rapidly declines below $f_{\mathrm{GW}}<100 \, \rm{Hz}$, so it may be difficult to detect slowly spinning stars.

We can now estimate the number of unknown NSs probed at a given $\epsilon$, defined as $N_\star$. We define our grid of $\epsilon$ values to range from $\log(\epsilon)=[-9,-5]$. Then, we perform the following integral to calculate $N_\star$
\begin{eqnarray}
\label{eq:N-star}
N_\star(\epsilon) = \int_{\mathrm \log(f_1)}^{\mathrm \log(f_2)} N(d(\log f,\epsilon))\times\\
\Phi(\log f)\mathrm{d}(\log f)\nonumber
\end{eqnarray}
where $f_1=20\, \mathrm{Hz}$ is the minimum sensitivity of GW detectors and $f_2$ is the spin-down limited frequency that a NS could produce a detectable signal in GWs. This value is dependent on the particular search and will mostly affect the number of fast-spinning highly elliptical NSs. In \cite{all-sky:2019}, the maximum spin-down considered in their search was $|\dot{f}|=1\times10^{-8}\, \mathrm{Hz \ s^{-1}}$ for the \textit{SkyHough} and \textit{TDFStat} pipelines and $|\dot{f}|=2\times10^{-9}\, \mathrm{Hz \ s^{-1}}$ for \textit{Frequency-Hough}. It should be noted that the limit on $|\dot{f}|$ for Frequency-Hough quoted here is the value considered in the range of 512-1024 Hz, whereas in the range of 20-512 Hz $|\dot{f}|=1\times10^{-8}\, \mathrm{Hz \ s^{-1}}$. For a NS with ellipticity $\epsilon$ 
\begin{eqnarray}
\label{eq:fmax}
f_{\mathrm{max}} = 225 \, \rm{ Hz} \, \Big(\frac{|\dot{\it{f}}|^{1/5}}{\epsilon^{2/5}}\Big)
\end{eqnarray} where $|\dot{f}|$ is in units of $[\mathrm{Hz \ s^{-1}}]$ \citep{spin-down}. Using the strain data from \cite{all-sky:2019}, we will adopt $f_2=f_{\rm max}$ for $|\dot{f}|=2\times10^{-9}\, \mathrm{Hz \ s^{-1}}$ for simplicity since the overall search is limited by the smallest maximum value of $|\dot{f}|$.

\section{Results}\label{sec:results}
\subsection{Neutron Star Estimates}
\label{sec:estimates}
\begin{figure*}[htb]
    \centering
    \includegraphics[width=0.8\textwidth]{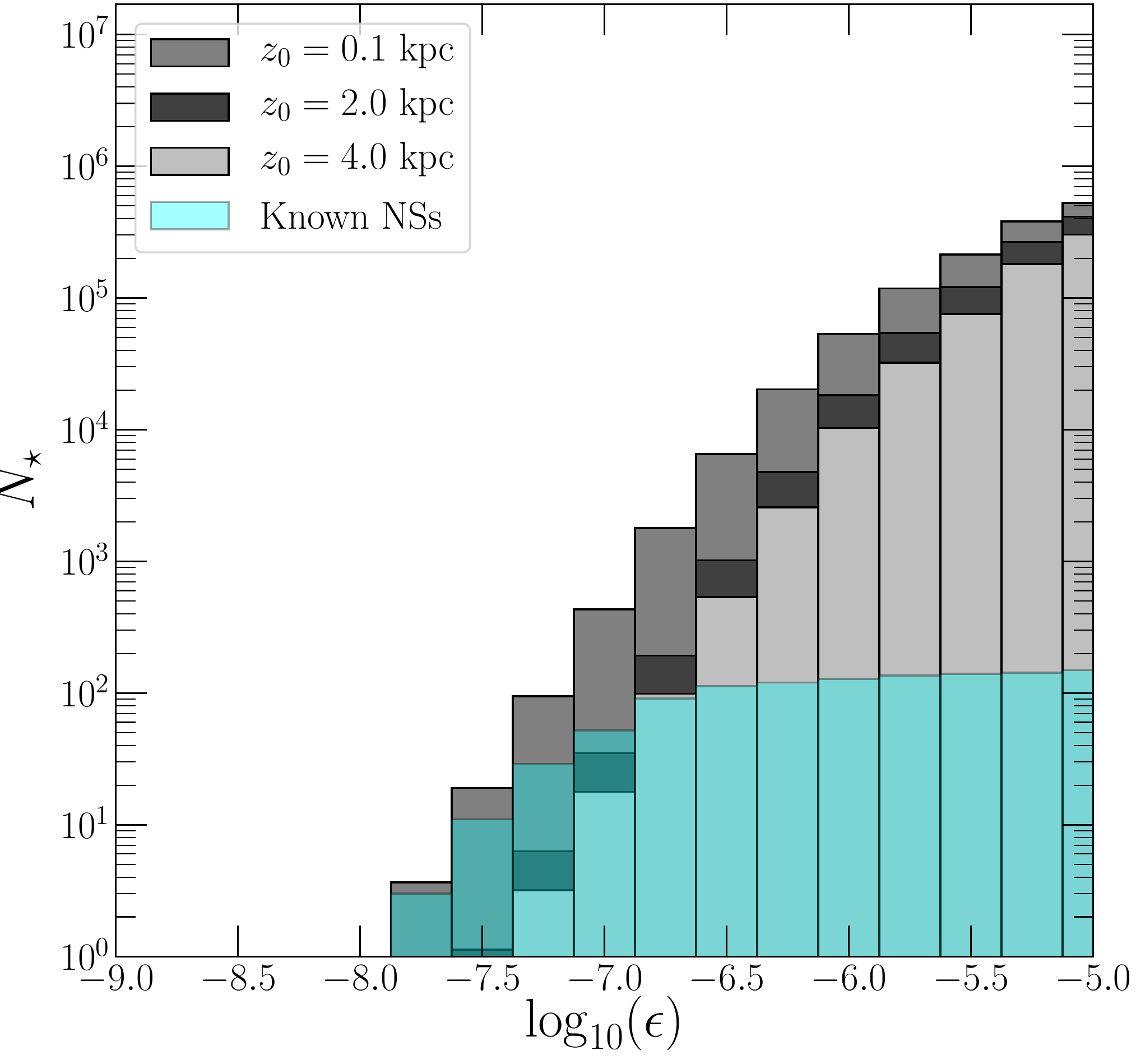}
    \caption{Predicted number of NSs probed by GW detectors to a given NS ellipticity. We show the predictions of \autoref{eq:N-star} for $z_0=0.1, 2.0,$ and $4.0\, \textrm{kpc}$. Also plotted are the limits on \textit{known} NS ellipticities derived from \cite{First_CGW} in light blue.}
    \label{fig:N-ell}
\end{figure*}

In \autoref{fig:N-ell} are model populations from solving \autoref{eq:N-star} for the chosen values of $z_0$ and these show the number of NSs probed to a given ellipticity. We also record the characteristic values for each of the models in \autoref{fig:N-ell} in \autoref{tab:eps_N}. We find that between $\approx 3\textrm{--}5.3\times10^5$ stars of the $1.7 \times 10^7$ NSs with $f_{\mathrm{GW}} > 20\, \mathrm{Hz}$ have been probed to $\epsilon\sim 10^{-5}$\textrm{---} corresponding to between $\approx 1.7\textrm{--}3.1\%$ of the model population. By contrast, only between $\approx 1\textrm{--}5.3\times10^4$ stars are probed to $\epsilon\sim 10^{-6}$, corresponding to only between $\approx 0.06\textrm{--}0.31\%$ of the visible population. For comparison, we show the limits on $\epsilon$ from the analysis of \cite{First_CGW} in \autoref{fig:N-ell} using CGW searches of \textit{known} pulsars at known $f_{\mathrm{GW}}$. This data probes significantly fewer NSs at ellipticities above $\epsilon \gtrsim 10^{-7}$ than using our method.  This is because most of the NS in the galaxy are unknown.  We discuss this further in \autoref{sec:discussion}.  

\begin{deluxetable}{c||c||c||c}[htb!]
\tabletypesize{\footnotesize}
\tablehead{$\log_{10}(\epsilon)$ & $z_0=0.1$ kpc & $z_0=2.0$ kpc & $z_0=4.0$ kpc}
\tablecaption{Estimates for number of NSs probed at the value of ellipticity $\epsilon$ for different values of disk thickness $z_0$.\label{tab:eps_N}}
\startdata
-5.00 & 5.3\tttt{5} & 4.1\tttt{5} & 3.0\tttt{5} \\
-5.25 & 3.8\tttt{5} & 2.7\tttt{5} & 1.8\tttt{5} \\
-5.50 & 2.1\tttt{5} & 1.2\tttt{5} & 7.6\tttt{4} \\
-5.75 & 1.2\tttt{5} & 5.4\tttt{4} & 3.2\tttt{4} \\
-6.00 & 5.3\tttt{4} & 1.8\tttt{4} & 1.0\tttt{4} \\
-6.25 & 2.0\tttt{4} & 4.8\tttt{3} & 2.6\tttt{3} \\
-6.50 & 6.5\tttt{3} & 1.0\tttt{3} & 540 \\
-6.75 & 1.8\tttt{3} & 190 & 99 \\
-7.00 & 430 & 35 & 18 \\
-7.25 & 95 & 6 & 3 \\
-7.50 & 19 & 1 & 1 \\
-7.75 & 4 & 0 & 0 \\
-8.00 & 1 & 0 & 0 \\
-8.25 & 0 & 0 & 0 \\
-8.50 & 0 & 0 & 0 \\
-8.75 & 0 & 0 & 0 \\
-9.00 & 0 & 0 & 0 \\
\enddata

\end{deluxetable}
\subsection{Effects of Improved Strain Sensitivity}
\label{sec:imp-sens}
The strain amplitude $h_0$ used in this study is limited by the sensitivity of GW interferometers and the parameters of the search. We can see the effects of improving the sensitivity of $h_0$ directly in \autoref{eq:dist} in that the distance we can be sensitive to will increase with decreasing $h_0$. This will then increase our estimate for the total number of NSs probed at a given ellipticity. As an example of this effect, we test what would happen to $N_\star$ should a new search reduce $h_0$ in either the high-frequency ($f_{\rm GW}\geq 1000 \rm{Hz}$) or low-frequency ($f_{\rm GW}\leq 100 \rm{Hz}$) regimes by a factor of two. 

\begin{figure*}[tb]
    \centering
    \includegraphics[width=0.8\textwidth]{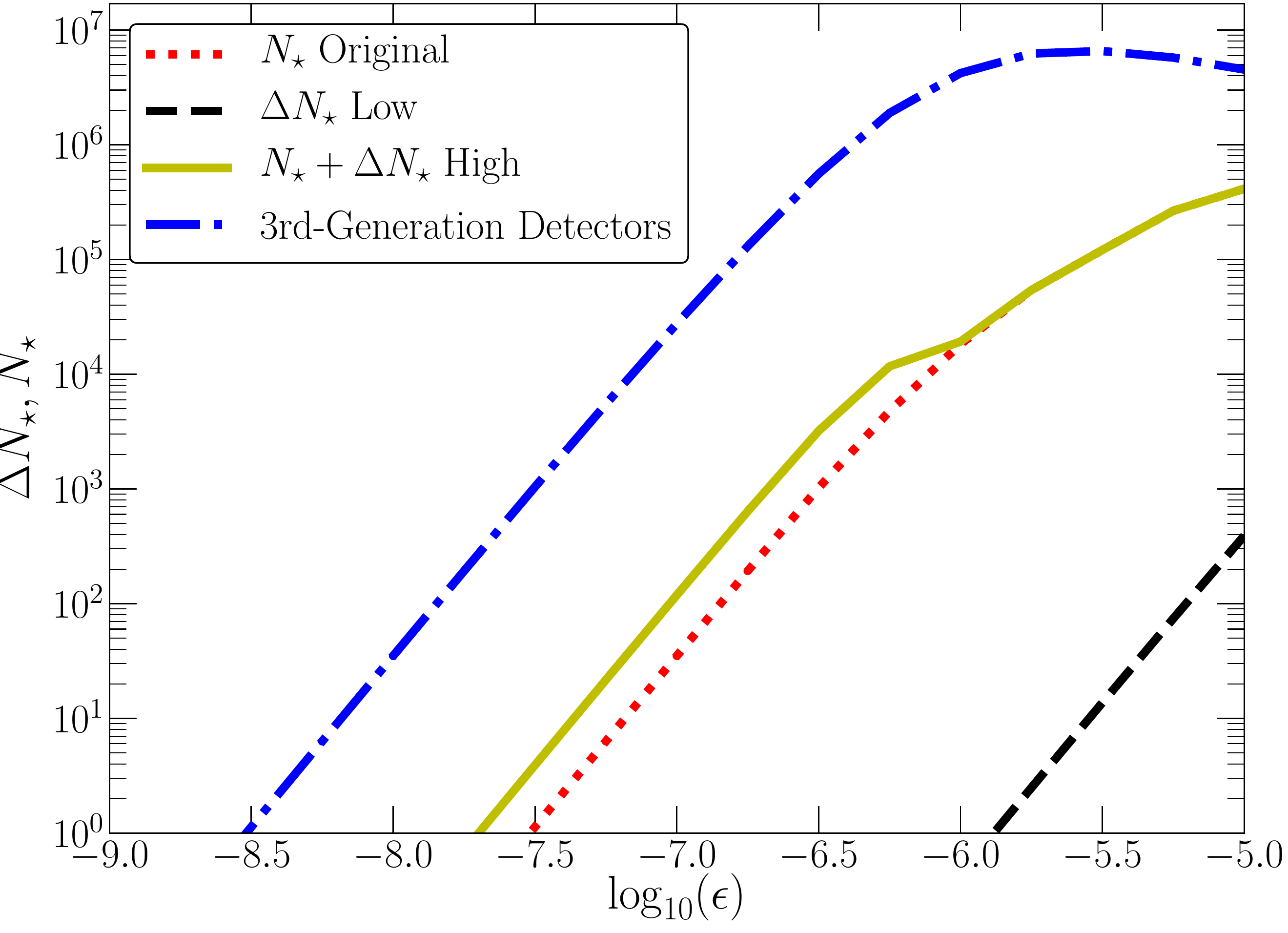}
    \caption{Number of additional NSs probed to a given ellipticity if the the strain sensitivity is improved using models with $z_0=2 \, \mathrm{kpc}$. The solid yellow line is the total number of new NSs when the strain amplitude is improved by a factor of two in the high frequency regime [$\geq 1000 \, \rm{Hz}$] and the dashed black line the effect of improving the low frequency regime [$\leq 500 \, \rm{Hz}$] by a factor of two. The blue dot-dash line is the $N_\star$ for a 10 times better strain sensitivity potentially achievable with the third generation of GW detectors. The dotted red line represents the original data for $z_0=2$ kpc from \autoref{fig:N-ell}.}
    \label{fig:deltaN}
\end{figure*}

We present the predicted values of $N_\star$ for improved detector sensitivity in \autoref{fig:deltaN} for a disk model which has $z_0=2.0\, \mathrm{kpc}$ and we also tabulate characteristic values in \autoref{tab:eps_N2}. In this figure, we show the original prediction for $N_\star$ in \autoref{fig:N-ell} along with the number of \textit{new} NSs probed, $\Delta N_\star$, when $h_0$ is decreased by a factor of two in the high or low frequency regimes. We find that improving the high-frequency regime by a factor of two has a much larger effect on the number estimates compared to improving the low-frequency by a factor of two. This is due to the larger number of MS-pulsars from our catalog compared to those with $f_{\mathrm GW}\lesssim 100$ Hz and because one is sensitive to greater distances at higher frequencies. In this simple example, we see that lowering the value of $h_0$ by a factor of two in the high frequency regime can add nearly three times as many \textit{new} NSs as the current estimates for $\epsilon\lesssim 10^{-6}$. 

While improved sensitivity in the high-frequency regime will increase $N_\star$, it's also worth examining the sensitivity of future third-generation GW detectors \textrm{---} for example the \emph{Einstein Telescope} \citep{punturo:2010} and \emph{Cosmic Explorer} \citep{dwyer:2015}. This generation of detectors at present is estimated to be a factor of ten times more sensitive than present detectors. To explore this possibility we revisit the model of $N_\star$ with $z_0=2 \, \mathrm{kpc}$, but with $h_0$ reduced by a factor of ten at \textit{all} frequencies. 

The resulting improvement to $N_\star$ \textrm{--} which we define as $\Delta N_{3g}$ \textrm{--} is shown in \autoref{fig:deltaN}. We see that there is an increase in $N_\star$ for all $\epsilon$ by at least an order-of-magnitude and for $\epsilon \sim 10^{-7}$ the improvement is approximately three orders-of-magnitude. While for the original study we were well below the observational limit of $\sim 17$ million NSs (probing $\lesssim 3 \, \%$), with the sensitivity of third generation GW detectors this limit is much closer to being reached (probing $\lesssim 40 \, \%$), see \autoref{tab:eps_N2}. We note that this assumes the same search parameters as in \cite{all-sky:2019}. An improved search could further improve these limits as well.

\begin{deluxetable}{c||c||c||c}
\caption{Estimates for number of new NSs probed with $z_0=2.0$ kpc when the strain amplitude $h_0$ sensitivity in different frequency regimes is increased by a factor of 2. The rightmost column is the result of decreasing the strain at all frequencies by a factor of 10.\label{tab:eps_N2}}
\tablehead{$\log_{10}(\epsilon)$ & $\Delta N_\star$($\leq 100$ \rm{Hz}) & $\Delta N_\star$($\geq 1000$ \rm{Hz}) & $\Delta N_{3g}$}
\startdata
-5.00 & 390 & 0 & 4.1\tttt{6} \\
-5.25 & 74 & 0 & 5.5\tttt{6} \\
-5.50 & 14 & 0 & 6.4\tttt{6} \\
-5.75 & 3 & 0 & 6.2\tttt{6} \\
-6.00 & 0 & 920 & 4.2\tttt{6} \\
-6.25 & 0 & 6.9\tttt{3} & 1.9\tttt{6} \\
-6.50 & 0 & 2.2\tttt{3} & 5.6\tttt{5} \\
-6.75 & 0 & 450 & 1.3\tttt{5} \\
-7.00 & 0 & 84 & 2.8\tttt{4} \\
-7.25 & 0 & 15 & 5.5\tttt{3} \\
-7.50 & 0 & 3 & 1.0\tttt{3} \\
-7.75 & 0 & 1 & 190 \\
-8.00 & 0 & 0 & 35 \\
-8.25 & 0 & 0 & 6 \\
-8.50 & 0 & 0 & 1 \\
-8.75 & 0 & 0 & 0 \\
-9.00 & 0 & 0 & 0 \\
\enddata
\end{deluxetable}

\subsection{Alternative Searches}
\label{sec:application}
\begin{figure}[htb]
    \centering
    \includegraphics[width=\columnwidth]{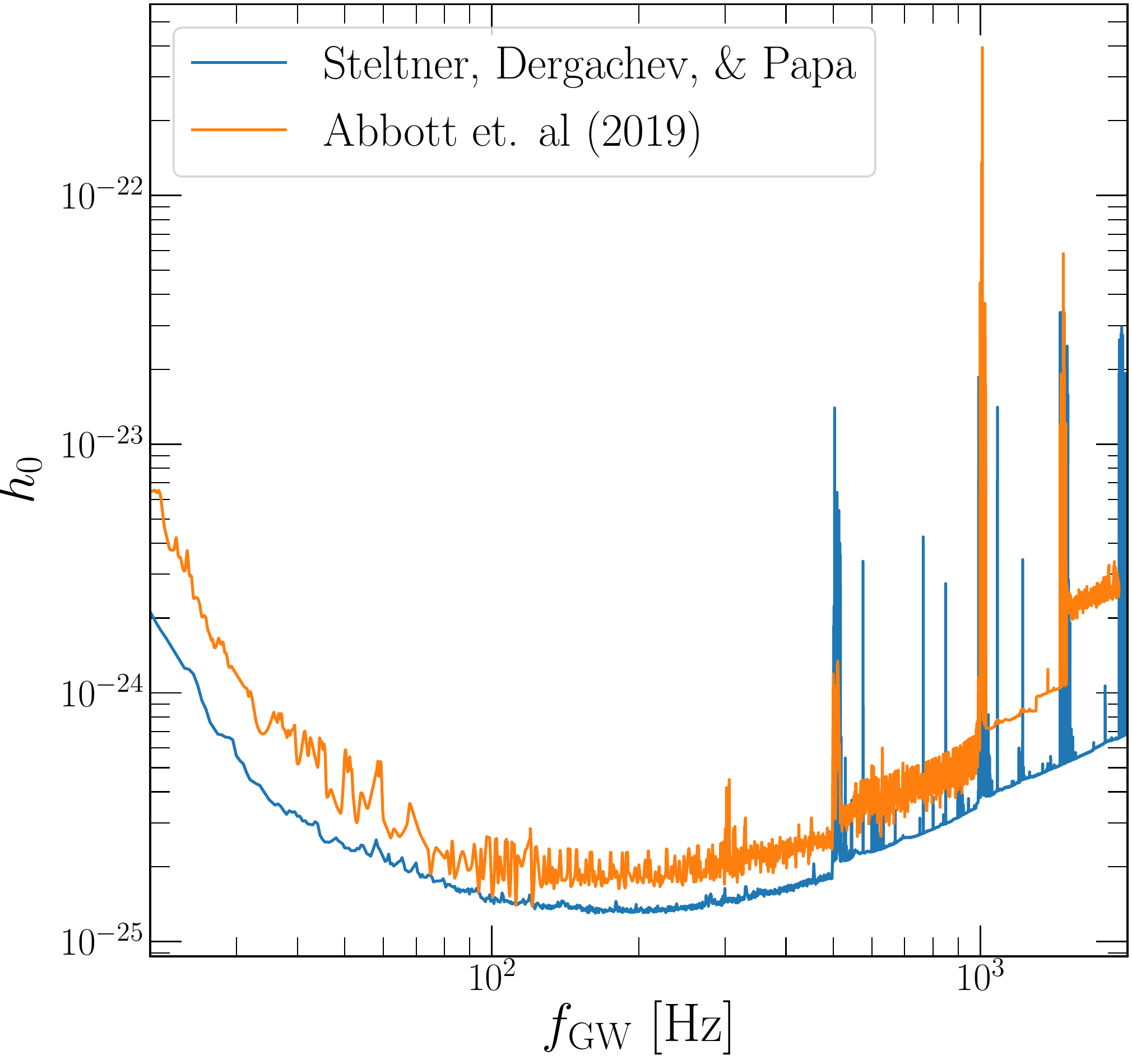}
    \caption{Strain sensitivity used in this study. We show in the blue the strain data described in \autoref{sec:application} and the data used in \autoref{sec:estimates} in orange. In addition to the noise being much less throughout, the blue data also has a smaller $h_0$ for the majority of frequencies.}
    \label{fig:map_data}
\end{figure}

We present now an example of our methodology using new data for the strain sensitivity. Here, we follow the same process for both the determination of $\Phi(f)$ and $N(d)$ as described in \autoref{sec:modelling}. For our strain sensitivity, we use the results of \cite{steltner2020einsteinhome} for frequencies 20 Hz - 500 Hz, \cite{dergachev:2020} for frequencies 500 Hz - 1700 Hz, and \cite{Dergachev_Papa:2021a} for frequencies 1700 Hz - 2000 Hz. We show this data in \autoref{fig:map_data}. These latter two searches were intended to search for low ellipticity NSs. As such, the maximum spin-down allowed for a given ellipticity is considerably lower, $|\dot{f}|=2.5\times10^{-12}\, \mathrm{Hz \ s^{-1}}$. The search performed by \cite{steltner2020einsteinhome} did consider a higher spin-down, $|\dot{f}|=2.6\times10^{-9}\, \mathrm{Hz \,  s^{-1}}$. For this reason, we have now split up the integral in \autoref{eq:N-star} with one integral being $f_1=20\, \mathrm{Hz}\, \& \, f_2=f_\mathrm{max}(|\dot{f}|=2.6$\tttt{-9}$\, \mathrm{Hz s}^{-1})$ and the second being $f_1=500\, \mathrm{Hz}\, \& \, f_2=f_\mathrm{max}(|\dot{f}|=2.5$\tttt{-12}$\, \mathrm{Hz s}^{-1})$. This ensures that we do not hinder the breadth of the search of \cite{steltner2020einsteinhome} with the lower $|\dot{f}|$ value.

\begin{figure}[htb]
    \centering
    \includegraphics[width=\columnwidth]{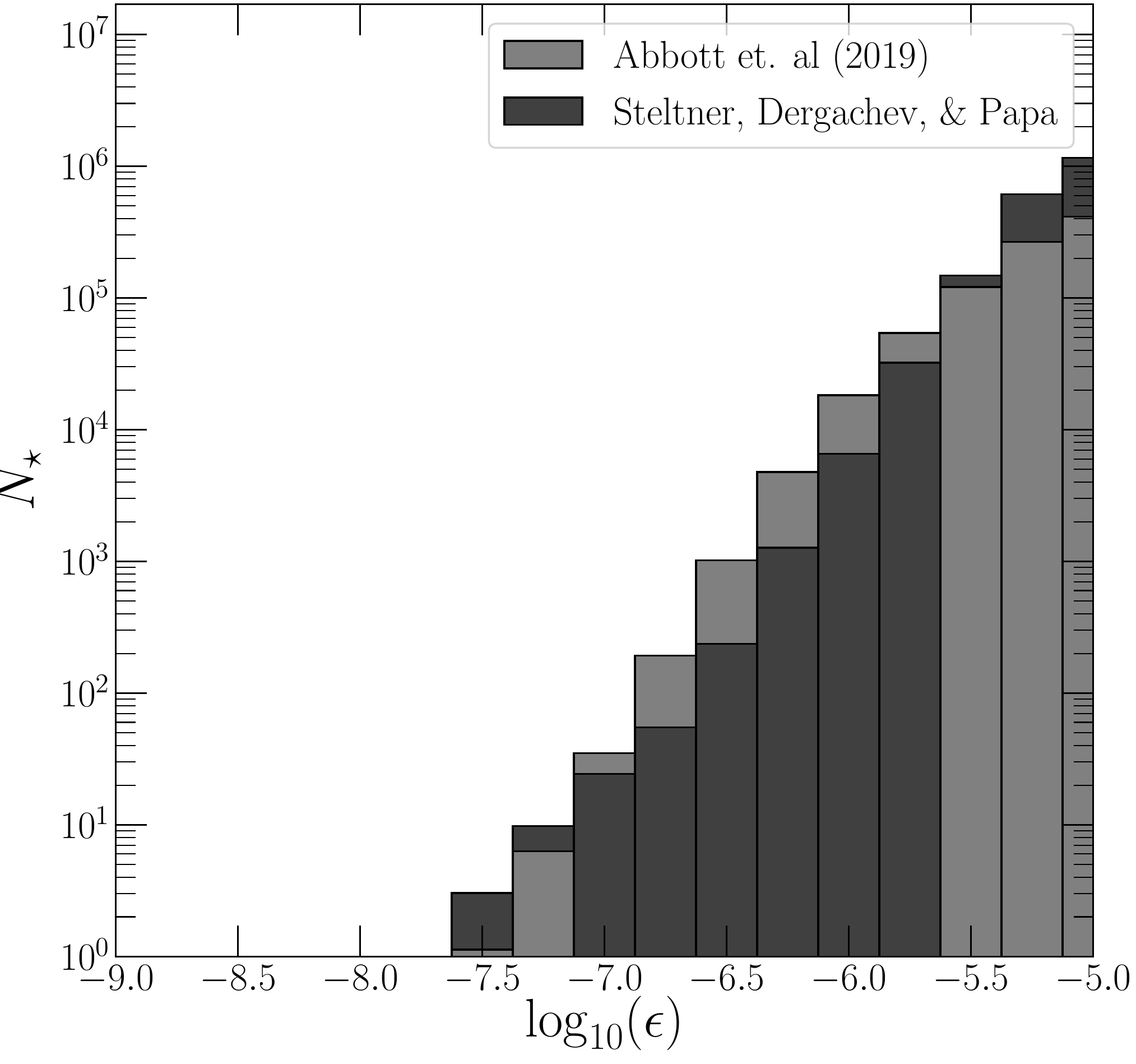}
    \caption{Models for $N_\star$ with $z_0=2$ kpc using strain data from \cite{all-sky:2019} (light) or \cite{steltner2020einsteinhome}, \cite{dergachev:2020}, and \cite{dergachev:2020} (dark).}
    \label{fig:map_res}
\end{figure}

Indeed, our results show that using this improved data does yield more NSs probed at small $\epsilon$, see \autoref{fig:map_res}. This can largely be attributed to the overall decrease in $h_0$ at $f_{\rm GW}\geq500$ Hz when compared to the strain from \cite{all-sky:2019}. However, at $\epsilon\gtrsim10^{-7}$ this data set probes significantly fewer NSs than for \cite{all-sky:2019}. This is expected since NSs with these high $\epsilon$ were not prioritized in the study of $h_0(f_\mathrm{GW})$ cited above. We note that the low-frequency regime has been considered by \cite{Dergachev_Papa:2021b} since this work was submitted.


\section{Discussion}\label{sec:discussion}
Prior searches for CGWs from {\it known} pulsars involve searching a well-defined number of NSs near an expected \fgw \ for each source \citep{Abbott:2017}. Limits on $\epsilon$ in \autoref{fig:N-ell} show the number of \textit{unknown} NSs where GW (assuming a source with a given $\epsilon$) have been searched for and not found. We see that the models begin to result in similar estimates near $N_\star\sim 10^{6}$ above $\epsilon\gtrsim10^{-6}$. This may be the maximum $\epsilon$ allowed by the NS's crust \citep{ushomirsky:2000,PhysRevLett.102.191102,10.1093/mnras/staa3635}, which is only slightly disfavored by our results. We predict that only $\gtrsim10^5$, or 0.1\%, of Galactic NSs have been probed above $\epsilon=10^{-5.5}$. This puts a limit on about one in ten million NSs may have such an ellipticity or we would have detected a signal in gravitational waves.

The largest ellipticity in our tested range $\epsilon = 10^{-5}$, though heavily disfavored from studies of the breaking strain of a NS's crust, cannot be ruled out entirely using current CGW data. From our results, we only rule out this ellipticity for $\approx 1.6\%$ of all Galactic NSs. This may be somewhat unrealistic, for example, a millisecond pulsar would produce a very large strain amplitude in CGW signal with such a large ellipticity. 

The theoretical upper limit on $\epsilon$ of $\sim\textrm{few }\times10^{-6}$ has likewise not been ruled out. In fact, our results suggest that $\lesssim 0.1\%$ of all Galactic NSs have been probed at this ellipticity for all values of the disk thickness. Therefore there is great need to continue searching for CGWs arising from NSs with ellipticities near this value. With further studies of NS ellipticities and CGW searches, this limit may become more apparent.

Our methodology used in this study attempted to keep things as simple as possible. Several additional complications to the study could be introduced to further constrain $N_\star$. Firstly, our choice of \autoref{eq:dist} as the distribution of Galactic NSs is a simple model which largely follows the star formation pattern in the Galactic disk. In reality, NSs may have a much different distribution, in part resulting from large transverse velocity kicks during their birth. We have attempted to mitigate this by picking different values for $z_0$ which either condense ($z_0=0.1\, \mathrm{kpc}$) or expand ($z_0=4.0\, \mathrm{kpc}$) the density distribution of NSs as seen from Earth. 

Most NSs are expected to be born with a transverse space-velocity from a supernovae kick \citep{shklovskii:1970}. Analysis of NS orbits suggests that fewer than $\lesssim 20\,\%$ are retained in the disk and a greater fraction remain in bound orbits in the Galactic Halo~\citep{sartore:2010}. Furthermore, some NSs have a sufficient space-velocity to escape the Galactic potential entirely ~\citep{arzoumanian:2002,katsuda:2018,nakamura:2019}. As a result, kicked NSs leaving the disk will spread the density distribution in \autoref{3d_dist} to larger $z_0$ than is typical for other stellar populations. Additionally, our calculated estimate for the total number of NSs probed in \autoref{tab:eps_N} could be reduced by more than factor of two depending on the real distribution of supernovae kick velocities. A clear next step with this type of estimate would be to self-consistently include an empirical density distribution of NSs that can account for a kicked NS population. 

Additionally, we have chosen to neglect CGWs arising from NSs in binaries because of the added complication it would cause on the GW signal and on the search parameters. However, in future work, it would be useful if the GW search treated binary NSs and isolated NSs separately. The newly developed \textit{BinarySkyHough} \citep{Covas:2019} pipeline is much better equipped to search for CGWs in binaries than its predecessor \textit{SkyHough} \citep{Covas:2004} which was used in \citet{all-sky:2019}. By better constraining the values of $\epsilon$, this can also further improve the search parameter computation time. 

We find that the disk thickness parameter $z_0$ from \autoref{3d_dist} has a significant impact on the estimated number of nearby NS. These nearby sources of CGWs would be vital in constraining ellipticities $\lesssim 10^{-7}$. In the thin disk approximation ($z_0\longrightarrow0$), \autoref{eq:dist} then goes like $\rho(d)\approx d$ for small values of $d$. However, for other values of $z_0$ where $d\ll z_0$, the distribution instead goes like $\rho(d)\approx d^2$. This has significant effects on nearby number estimates as any stars lying above the plane of the disk are then condensed, thereby increasing the total number of stars estimated. This is easily seen in the right-hand figure of \autoref{num_dens}, whose effects on the estimated numbers of NSs seen in \autoref{fig:N-ell} and in \autoref{tab:eps_N}. \autoref{fig:N-ell} is the result of \autoref{eq:N-star} for the values of $z_0$ used in \autoref{tab:constants}. We see that $N_\star$ is very sensitive to $z_0$ for small values of $\epsilon$, decreasing for increasing $z_0$.  Better determination of the disk thickness of the Galactic NS population is important for constraining $\epsilon\lesssim10^{-5.5}$, where an order of magnitude difference exists between the models used here. 

We explore the implications of a new CGW search with improved $h_0(f_{\rm GW})$ sensitivity using the current generation of GW detectors. Our results show that improving sensitivity in the high-frequency regime ($f_{\rm GW} \geq 1000 \, \mathrm{Hz}$) can have the greatest impact on the search for CGWs. From \autoref{fig:deltaN}, we can see that the improvements in $h_0$ sensitivity can have much higher returns on the total number of new NSs probed. At $\epsilon\lesssim10^{-6}$, for example, this results in finding approximately \textit{three} times $N_\star$ new NSs.

Interestingly, this is not true for very high values of $\epsilon$. We see that the high frequency regime has a turnoff point at $\epsilon\sim 10^{-6}$ which occurs for two reasons. First, the original search already probed a significant fraction of visible NSs in the disk for $\epsilon>10^{-6}$, and so fewer new NSs would become visible. Secondly, for $\epsilon>10^{-5.5}$, $f_{\mathrm{max}}<1000 \, \mathrm{Hz}$ and so the improvement is no longer limiting the contribution of $h_0(f_\mathrm{GW})$ to the integral in \autoref{eq:N-star}. From these two points, we see that the strain sensitivity at high frequency has a significant impact on searches for NSs with moderate ellipticity. Conversely, improving the low-frequency regime ($f_{\rm GW} \leq 100 \, \mathrm{Hz}$) certainly increases the number probed, it is approximately \textit{four} orders of magnitude smaller in effect than improving the high-frequency regime for moderate ellipticity. This is because there are a much larger number of MS-pulsars with spin frequencies in excess of $\nu > 50$ Hz, as discussed in \autoref{sec:spin}.

Third generation detectors may dramatically increase the number of NSs probed. Given the blue dash-dot curve in \autoref{fig:deltaN} we can see that improving $h_0$ by a factor of ten increases $N_{\star}$ by more than a factor of \textit{100-1000 times}. We note that this assumes for \autoref{eq:fmax} $|\dot{f}|=2\times10^{-9} \, \mathrm{Hz\, s^{-1}}$ which may not be the true limit considered when searches with these instruments take place. Despite this, however, just the improvements to $h_0$ we estimate will probe almost 40\% of all the NSs in the Galaxy at large $\epsilon$. In addition, improvements in search techniques and computer resources may further increase the number of NSs probed.

We conclude our discussion with the analysis of \autoref{sec:application}. The data used here is comprised of several additional analyses of the data from \cite{all-sky:2019}, however now with improved strain sensitivity. Both searches use different techniques and have different goals for performing their respective searches. For example, \citet{dergachev:2020} were primarily interested in finding low $\epsilon$ neutron stars which allows for a smaller maximum value for $|\dot{f}|$. If one sets $f_{\mathrm{max}}=2000$ Hz, then for $\epsilon=10^{-8}$ $|\dot{f}|\approx5.55\times10^{-12}\, \mathrm{Hz\, s^{-1}}$. This allows for more restrictive limits on $h_0$, as seen in \autoref{fig:map_data}. While the value of $|\dot{f}|=2.5\times10^{-12} \, \mathrm{Hz\, s^{-1}}$ is consistent with pulsar data, this constraint on the analysis has two effects on the overall results. First, it reduces the search parameter space considerably and therefore allows for a better determination of $h_0(f_\mathrm{GW})$, as seen in \autoref{fig:map_data}. As we have shown in \autoref{sec:imp-sens}, reducing the strain amplitude does increase $N_\star$.

However, the second, and most important effect for this work, is that it limits the amount of detectable NSs. For example, taking $|\dot{f}|=2\times10^{-9}\, \mathrm{Hz \ s^{-1}}$ as we did in \autoref{sec:estimates}, for $\epsilon=10^{-6}$ this means $f_{\rm max}\approx1029 \, \mathrm{Hz}$. Using $|\dot{f}|=2.5\times10^{-12}\, \mathrm{Hz \ s^{-1}}$ instead, $f_{\rm max}\approx220 \, \mathrm{Hz}$. This means that this data set may be inefficient when looking for highly elliptical neutron stars because none of the MSPs are being probed. Interestingly though, the improved data set does probe more NSs at $\epsilon>10^{-5.75}$. This is because the values of $f_\mathrm{max}$ for both sets exclude the highest frequencies from the search, meaning only the low frequency sources contribute to $N_\star$. Since we see in \autoref{fig:map_data} that the strain used in this analysis is smaller than from \cite{all-sky:2019}, slightly more NSs are probed.

Note that there are two possible approaches when selecting the optimal choice for $|\dot{f}|$ because the ellipticity distribution of NSs is unknown. Should a future CGW search occur with the intent of probing the highest ellipticities near $\epsilon \sim 10^{-6}$ one should consider using a higher $|\dot{f}|$ limit. Possibly the most promising value of $|\dot{f}|$ is slightly higher than considered by \cite{all-sky:2019}, $5.55\times 10^{-8}\, \mathrm{Hz} \ s^{-1}$. This value will ensure that $f_\mathrm{max}=2000$ Hz for $\epsilon=10^{-6}$ so that any fast-spinning NSs won't be excluded. On the other hand, if the intent is to find lower ellipticity NSs $\textrm{--}$ for instance, near $\epsilon \sim 10^{-9}$ $\textrm{--}$ one should consider a deeper search with a lower $|\dot{f}|$ limit.

\section{Conclusion}
We have detailed estimates on the total number of NSs probed with gravitational wave detectors. In doing so, we have shown that continuous gravitational wave searches suggest that fewer than about one in ten thousand NSs have an ellipticity $\gtrsim10^{-6}$. Additionally, we have shown that the disk thickness strongly affects the number counts of nearby neutron stars while leaving more distant stars largely unaffected. We have explored the effects of improving strain amplitude sensitivity at higher frequencies which can increase the amount of NSs probed to a given ellipticity. These estimates are important for setting upper limits on the ellipticity of a NS as well as detecting radio quiet neutron stars that may be nearby, yet unobserved. Finally, we discuss the impact of third-generation detectors and find that they may probe 100-1000 times more NSs than have presently been probed.

\begin{acknowledgments}
We would like to thank the anonymous referee for their helpful comments and kind words. We would also like to extend our gratitude to M. Papa for useful comments and for providing the data used in \autoref{sec:application}. We also thank N. Andersson, F. Gittins, V. Dergachev, F. De Lillo, and P. Covas for their  helpful comments.

This material is based upon work supported by the U.S. Department of Energy Office of Science, Office of Nuclear Physics under Awards DE-FG02-87ER40365 (Indiana University) and Number DE-SC0008808 (NUCLEI SciDAC Collaboration). 
\end{acknowledgments}

\bibliographystyle{yahapj}
\bibliography{references.bib}

\begin{thebibliography}{}
\providecommand\natexlab[1]{#1}
\providecommand\JournalTitle[1]{#1}

\bibitem[{{Abbott} {et~al.}(2017){Abbott}, {Abbott}, {Abbott}, {Abernathy},
  {Acernese}, {Ackley}, {Adams}, {Adams}, {Addesso}, {Adhikari}, {Adya},
  {Affeldt}, {Agathos}, {Agatsuma}, {Aggarwal}, {Aguiar}, {Aiello}, {Ain},
  {Ajith}, {Allen}, {Allocca}, {Altin}, {Ananyeva}, {Anderson}, {Anderson},
  {Appert}, {Arai}, {Araya}, {Areeda}, {Arnaud}, {Arun}, {Ascenzi}, {Ashton},
  {Ast}, {Aston}, {Astone}, {Aufmuth}, {Aulbert}, {Avila-Alvarez}, {Babak},
  {Bacon}, {Bader}, {Baker}, {Baldaccini}, {Ballardin}, {Ballmer}, {Barayoga},
  {Barclay}, {Barish}, {Barker}, {Barone}, {Barr}, {Barsotti}, {Barsuglia},
  {Barta}, {Bartlett}, {Bartos}, {Bassiri}, {Basti}, {Batch}, {Baune},
  {Bavigadda}, {Bazzan}, {Beer}, {Bejger}, {Belahcene}, {Belgin}, {Bell},
  {Berger}, {Bergmann}, {Berry}, {Bersanetti}, {Bertolini}, {Betzwieser},
  {Bhagwat}, {Bhandare}, {Bilenko}, {Billingsley}, {Billman}, {Birch},
  {Birney}, {Birnholtz}, {Biscans}, {Bisht}, {Bitossi}, {Biwer}, {Bizouard},
  {Blackburn}, {Blackman}, {Blair}, {Blair}, {Blair}, {Bloemen}, {Bock},
  {Boer}, {Bogaert}, {Bohe}, {Bondu}, {Bonnand}, {Boom}, {Bork}, {Boschi},
  {Bose}, {Bouffanais}, {Bozzi}, {Bradaschia}, {Brady}, {Braginsky},
  {Branchesi}, {Brau}, {Briant}, {Brillet}, {Brinkmann}, {Brisson}, {Brockill},
  {Broida}, {Brooks}, {Brown}, {Brown}, {Brown}, {Brunett}, {Buchanan},
  {Buikema}, {Bulik}, {Bulten}, {Buonanno}, {Buskulic}, {Buy}, {Byer},
  {Cabero}, {Cadonati}, {Cagnoli}, {Cahillane}, {Calder{\'o}n Bustillo},
  {Callister}, {Calloni}, {Camp}, {Canepa}, {Cannon}, {Cao}, {Cao}, {Capano},
  {Capocasa}, {Carbognani}, {Caride}, {Casanueva Diaz}, {Casentini}, {Caudill},
  {Cavagli{\`a}}, {Cavalier}, {Cavalieri}, {Cella}, {Cepeda}, {Cerboni
  Baiardi}, {Cerretani}, {Cesarini}, {Chamberlin}, {Chan}, {Chao}, {Charlton},
  {Chassande-Mottin}, {Cheeseboro}, {Chen}, {Chen}, {Cheng}, {Chincarini},
  {Chiummo}, {Chmiel}, {Cho}, {Cho}, {Chow}, {Christensen}, {Chu}, {Chua},
  {Chua}, {Chung}, {Ciani}, {Clara}, {Clark}, {Cleva}, {Cocchieri}, {Coccia},
  {Cohadon}, {Colla}, {Collette}, {Cominsky}, {Constancio}, {Conti}, {Cooper},
  {Corbitt}, {Cornish}, {Corsi}, {Cortese}, {Costa}, {Coughlin}, {Coughlin},
  {Coulon}, {Countryman}, {Couvares}, {Covas}, {Cowan}, {Coward}, {Cowart},
  {Coyne}, {Coyne}, {Creighton}, {Creighton}, {Cripe}, {Crowder}, {Cullen},
  {Cumming}, {Cunningham}, {Cuoco}, {Dal Canton}, {Danilishin}, {D'Antonio},
  {Danzmann}, {Dasgupta}, {Da Silva Costa}, {Dattilo}, {Dave}, {Davier},
  {Davies}, {Davis}, {Daw}, {Day}, {Day}, {De}, {DeBra}, {Debreczeni},
  {Degallaix}, {De Laurentis}, {Del{\'e}glise}, {Del Pozzo}, {Denker}, {Dent},
  {Dergachev}, {De Rosa}, {DeRosa}, {DeSalvo}, {Devenson}, {Devine},
  {Dhurandhar}, {D{\'\i}az}, {Di Fiore}, {Di Giovanni}, {Di Girolamo}, {Di
  Lieto}, {Di Pace}, {Di Palma}, {Di Virgilio}, {Doctor}, {Dolique}, {Donovan},
  {Dooley}, {Doravari}, {Dorrington}, {Douglas}, {Dovale {\'A}lvarez},
  {Downes}, {Drago}, {Drever}, {Driggers}, {Du}, {Ducrot}, {Dwyer}, {Edo},
  {Edwards}, {Effler}, {Eggenstein}, {Ehrens}, {Eichholz}, {Eikenberry},
  {Eisenstein}, {Essick}, {Etienne}, {Etzel}, {Evans}, {Evans}, {Everett},
  {Factourovich}, {Fafone}, {Fair}, {Fairhurst}, {Fan}, {Farinon}, {Farr},
  {Farr}, {Fauchon-Jones}, {Favata}, {Fays}, {Fehrmann}, {Fejer},
  {Fern{\'a}ndez Galiana}, {Ferrante}, {Ferreira}, {Ferrini}, {Fidecaro},
  {Fiori}, {Fiorucci}, {Fisher}, {Flaminio}, {Fletcher}, {Fong}, {Forsyth},
  {Fournier}, {Frasca}, {Frasconi}, {Frei}, {Freise}, {Frey}, {Frey}, {Fries},
  {Fritschel}, {Frolov}, {Fulda}, {Fyffe}, {Gabbard}, {Gadre}, {Gaebel},
  {Gair}, {Gammaitoni}, {Gaonkar}, {Garufi}, {Gaur}, {Gayathri}, {Gehrels},
  {Gemme}, {Genin}, {Gennai}, {George}, {Gergely}, {Germain}, {Ghonge},
  {Ghosh}, {Ghosh}, {Ghosh}, {Giaime}, {Giardina}, {Giazotto}, {Gill},
  {Glaefke}, {Goetz}, {Goetz}, {Gondan}, {Gonz{\'a}lez}, {Gonzalez Castro},
  {Gopakumar}, {Gorodetsky}, {Gossan}, {Gosselin}, {Gouaty}, {Grado}, {Graef},
  {Granata}, {Grant}, {Gras}, {Gray}, {Greco}, {Green}, {Groot}, {Grote},
  {Grunewald}, {Guidi}, {Guo}, {Gupta}, {Gupta}, {Gushwa}, {Gustafson},
  {Gustafson}, {Hacker}, {Hall}, {Hall}, {Hammond}, {Haney}, {Hanke}, {Hanks},
  {Hanna}, {Hanson}, {Hardwick}, {Harms}, {Harry}, {Harry}, {Hart}, {Hartman},
  {Haster}, {Haughian}, {Healy}, {Heidmann}, {Heintze}, {Heitmann}, {Hello},
  {Hemming}, {Hendry}, {Heng}, {Hennig}, {Henry}, {Heptonstall}, {Heurs},
  {Hild}, {Hoak}, {Hofman}, {Holt}, {Holz}, {Hopkins}, {Hough}, {Houston},
  {Howell}, {Hu}, {Huerta}, {Huet}, {Hughey}, {Husa}, {Huttner}, {Huynh-Dinh},
  {Indik}, {Ingram}, {Inta}, {Isa}, {Isac}, {Isi}, {Isogai}, {Iyer}, {Izumi},
  {Jacqmin}, {Jani}, {Jaranowski}, {Jawahar}, {Jim{\'e}nez-Forteza}, {Johnson},
  {Jones}, {Jones}, {Jonker}, {Ju}, {Junker}, {Kalaghatgi}, {Kalogera},
  {Kandhasamy}, {Kang}, {Kanner}, {Karki}, {Karvinen}, {Kasprzack},
  {Katsavounidis}, {Katzman}, {Kaufer}, {Kaur}, {Kawabe}, {K{\'e}f{\'e}lian},
  {Keitel}, {Kelley}, {Kennedy}, {Key}, {Khalili}, {Khan}, {Khan}, {Khan},
  {Khazanov}, {Kijbunchoo}, {Kim}, {Kim}, {Kim}, {Kim}, {Kim}, {Kimbrell},
  {King}, {King}, {Kirchhoff}, {Kissel}, {Klein}, {Kleybolte}, {Klimenko},
  {Koch}, {Koehlenbeck}, {Koley}, {Kondrashov}, {Kontos}, {Korobko}, {Korth},
  {Kowalska}, {Kozak}, {Kr{\"a}mer}, {Kringel}, {Krishnan}, {Kr{\'o}lak},
  {Kuehn}, {Kumar}, {Kumar}, {Kuo}, {Kutynia}, {Lackey}, {Landry}, {Lang},
  {Lange}, {Lantz}, {Lanza}, {Lartaux-Vollard}, {Lasky}, {Laxen}, {Lazzarini},
  {Lazzaro}, {Leaci}, {Leavey}, {Lebigot}, {Lee}, {Lee}, {Lee}, {Lee},
  {Lehmann}, {Lenon}, {Leonardi}, {Leong}, {Leroy}, {Letendre}, {Levin}, {Li},
  {Libson}, {Littenberg}, {Liu}, {Lockerbie}, {Lombardi}, {London}, {Lord},
  {Lorenzini}, {Loriette}, {Lormand}, {Losurdo}, {Lough}, {Lousto}, {Lovelace},
  {L{\"u}ck}, {Lundgren}, {Lynch}, {Ma}, {Macfoy}, {Machenschalk}, {MacInnis},
  {Macleod}, {Maga{\~n}a-Sandoval}, {Majorana}, {Maksimovic}, {Malvezzi},
  {Man}, {Mandic}, {Mangano}, {Mansell}, {Manske}, {Mantovani}, {Marchesoni},
  {Marion}, {M{\'a}rka}, {M{\'a}rka}, {Markosyan}, {Maros}, {Martelli},
  {Martellini}, {Martin}, {Martynov}, {Mason}, {Masserot}, {Massinger},
  {Masso-Reid}, {Mastrogiovanni}, {Matichard}, {Matone}, {Mavalvala},
  {Mazumder}, {McCarthy}, {McClelland}, {McCormick}, {McGrath}, {McGuire},
  {McIntyre}, {McIver}, {McManus}, {McRae}, {McWilliams}, {Meacher}, {Meadors},
  {Meidam}, {Melatos}, {Mendell}, {Mendoza-Gandara}, {Mercer}, {Merilh},
  {Merzougui}, {Meshkov}, {Messenger}, {Messick}, {Metzdorff}, {Meyers},
  {Mezzani}, {Miao}, {Michel}, {Middleton}, {Mikhailov}, {Milano}, {Miller},
  {Miller}, {Miller}, {Miller}, {Millhouse}, {Minenkov}, {Ming}, {Mirshekari},
  {Mishra}, {Mitra}, {Mitrofanov}, {Mitselmakher}, {Mittleman}, {Moggi},
  {Mohan}, {Mohapatra}, {Montani}, {Moore}, {Moore}, {Moraru}, {Moreno},
  {Morriss}, {Mours}, {Mow-Lowry}, {Mueller}, {Muir}, {Mukherjee}, {Mukherjee},
  {Mukherjee}, {Mukund}, {Mullavey}, {Munch}, {Muniz}, {Murray}, {Mytidis},
  {Napier}, {Nardecchia}, {Naticchioni}, {Nelemans}, {Nelson}, {Neri}, {Nery},
  {Neunzert}, {Newport}, {Newton}, {Nguyen}, {Nielsen}, {Nissanke}, {Nitz},
  {Noack}, {Nocera}, {Nolting}, {Normandin}, {Nuttall}, {Oberling}, {Ochsner},
  {Oelker}, {Ogin}, {Oh}, {Oh}, {Ohme}, {Oliver}, {Oppermann}, {Oram},
  {O'Reilly}, {O'Shaughnessy}, {Ottaway}, {Overmier}, {Owen}, {Pace}, {Page},
  {Pai}, {Pai}, {Palamos}, {Palashov}, {Palomba}, {Pal-Singh}, {Pan}, {Pankow},
  {Pannarale}, {Pant}, {Paoletti}, {Paoli}, {Papa}, {Paris}, {Parker},
  {Pascucci}, {Pasqualetti}, {Passaquieti}, {Passuello}, {Patricelli},
  {Pearlstone}, {Pedraza}, {Pedurand}, {Pekowsky}, {Pele}, {Penn}, {Perez},
  {Perreca}, {Perri}, {Pfeiffer}, {Phelps}, {Piccinni}, {Pichot},
  {Piergiovanni}, {Pierro}, {Pillant}, {Pinard}, {Pinto}, {Pitkin}, {Poe},
  {Poggiani}, {Popolizio}, {Post}, {Powell}, {Prasad}, {Pratt}, {Predoi},
  {Prestegard}, {Prijatelj}, {Principe}, {Privitera}, {Prix}, {Prodi},
  {Prokhorov}, {Puncken}, {Punturo}, {Puppo}, {P{\"u}rrer}, {Qi}, {Qin}, {Qiu},
  {Quetschke}, {Quintero}, {Quitzow-James}, {Raab}, {Rabeling}, {Radkins},
  {Raffai}, {Raja}, {Rajan}, {Rakhmanov}, {Rapagnani}, {Raymond}, {Razzano},
  {Re}, {Read}, {Regimbau}, {Rei}, {Reid}, {Reitze}, {Rew}, {Reyes}, {Rhoades},
  {Ricci}, {Riles}, {Rizzo}, {Robertson}, {Robie}, {Robinet}, {Rocchi},
  {Rolland}, {Rollins}, {Roma}, {Romano}, {Romie}, {Rosi{\'n}ska}, {Rowan},
  {R{\"u}diger}, {Ruggi}, {Ryan}, {Sachdev}, {Sadecki}, {Sadeghian},
  {Sakellariadou}, {Salconi}, {Saleem}, {Salemi}, {Samajdar}, {Sammut},
  {Sampson}, {Sanchez}, {Sandberg}, {Sanders}, {Sassolas}, {Sathyaprakash},
  {Saulson}, {Sauter}, {Savage}, {Sawadsky}, {Schale}, {Scheuer}, {Schmidt},
  {Schmidt}, {Schmidt}, {Schnabel}, {Schofield}, {Sch{\"o}nbeck}, {Schreiber},
  {Schuette}, {Schutz}, {Schwalbe}, {Scott}, {Scott}, {Sellers}, {Sengupta},
  {Sentenac}, {Sequino}, {Sergeev}, {Setyawati}, {Shaddock}, {Shaffer},
  {Shahriar}, {Shapiro}, {Shawhan}, {Sheperd}, {Shoemaker}, {Shoemaker},
  {Siellez}, {Siemens}, {Sieniawska}, {Sigg}, {Silva}, {Singer}, {Singer},
  {Singh}, {Singh}, {Singhal}, {Sintes}, {Slagmolen}, {Smith}, {Smith},
  {Smith}, {Son}, {Sorazu}, {Sorrentino}, {Souradeep}, {Spencer}, {Srivastava},
  {Staley}, {Steinke}, {Steinlechner}, {Steinlechner}, {Steinmeyer},
  {Stephens}, {Stevenson}, {Stone}, {Strain}, {Straniero}, {Stratta},
  {Strigin}, {Sturani}, {Stuver}, {Summerscales}, {Sun}, {Sunil}, {Sutton},
  {Swinkels}, {Szczepa{\'n}czyk}, {Tacca}, {Talukder}, {Tanner}, {T{\'a}pai},
  {Taracchini}, {Taylor}, {Theeg}, {Thomas}, {Thomas}, {Thomas}, {Thorne},
  {Thrane}, {Tippens}, {Tiwari}, {Tiwari}, {Tokmakov}, {Toland}, {Tomlinson},
  {Tonelli}, {Tornasi}, {Torrie}, {T{\"o}yr{\"a}}, {Travasso}, {Traylor},
  {Trifir{\`o}}, {Trinastic}, {Tringali}, {Trozzo}, {Tse}, {Tso}, {Turconi},
  {Tuyenbayev}, {Ugolini}, {Unnikrishnan}, {Urban}, {Usman}, {Vahlbruch},
  {Vajente}, {Valdes}, {van Bakel}, {van Beuzekom}, {van den Brand}, {Van Den
  Broeck}, {Vander-Hyde}, {van der Schaaf}, {van Heijningen}, {van Veggel},
  {Vardaro}, {Varma}, {Vass}, {Vas{\'u}th}, {Vecchio}, {Vedovato}, {Veitch},
  {Veitch}, {Venkateswara}, {Venugopalan}, {Verkindt}, {Vetrano}, {Vicer{\'e}},
  {Viets}, {Vinciguerra}, {Vine}, {Vinet}, {Vitale}, {Vo}, {Vocca}, {Vorvick},
  {Voss}, {Vousden}, {Vyatchanin}, {Wade}, {Wade}, {Wade}, {Walker}, {Wallace},
  {Walsh}, {Wang}, {Wang}, {Wang}, {Wang}, {Ward}, {Warner}, {Was}, {Watchi},
  {Weaver}, {Wei}, {Weinert}, {Weinstein}, {Weiss}, {Wen}, {We{\ss}els},
  {Westphal}, {Wette}, {Whelan}, {Whiting}, {Whittle}, {Williams}, {Williams},
  {Williamson}, {Willis}, {Willke}, {Wimmer}, {Winkler}, {Wipf}, {Wittel},
  {Woan}, {Woehler}, {Worden}, {Wright}, {Wu}, {Wu}, {Yam}, {Yamamoto},
  {Yancey}, {Yap}, {Yu}, {Yu}, {Yvert}, {Zadro{\.z}ny}, {Zangrando}, {Zanolin},
  {Zendri}, {Zevin}, {Zhang}, {Zhang}, {Zhang}, {Zhang}, {Zhao}, {Zhou},
  {Zhou}, {Zhu}, {Zhu}, {Zucker}, {Zweizig}, {LIGO Scientific Collaboration},
  {Virgo Collaboration}, {Buchner}, {Cognard}, {Corongiu}, {Freire},
  {Guillemot}, {Hobbs}, {Kerr}, {Lyne}, {Possenti}, {Ridolfi}, {Shannon},
  {Stappers}, \& {Weltevrede}}]{First_CGW}
{Abbott}, B.~P., {Abbott}, R., {Abbott}, T.~D., {et~al.} 2017,
  \href{http://dx.doi.org/10.3847/1538-4357/aa677f}{\JournalTitle{\apj}, 839,
  12}

\bibitem[{Abbott {et~al.}(2017)}]{Abbott:2017}
Abbott, B.~P., {et~al.} 2017,
  \href{http://dx.doi.org/10.1103/PhysRevLett.119.161101}{\JournalTitle{\prl},
  119, 161101}

\bibitem[{{Abbott} {et~al.}(2019{\natexlab{a}}){Abbott}, {Abbott}, {Abbott},
  {Abraham}, {Acernese}, {Ackley}, {Adams}, {Adhikari}, {Adya}, {Affeldt},
  {Agathos}, {Agatsuma}, {Aggarwal}, {Aguiar}, {Aiello}, {Ain}, {Ajith},
  {Allen}, {Allocca}, {Aloy}, {Altin}, {Amato}, {Ananyeva}, {Anderson},
  {Anderson}, {Angelova}, {Antier}, {Appert}, {Arai}, {Araya}, {Areeda},
  {Ar{\`e}ne}, {Arnaud}, {Arun}, {Ascenzi}, {Ashton}, {Aston}, {Astone},
  {Aubin}, {Aufmuth}, {AultONeal}, {Austin}, {Avendano}, {Avila-Alvarez},
  {Babak}, {Bacon}, {Badaracco}, {Bader}, {Bae}, {Baker}, {Baldaccini},
  {Ballardin}, {Ballmer}, {Banagiri}, {Barayoga}, {Barclay}, {Barish},
  {Barker}, {Barkett}, {Barnum}, {Barone}, {Barr}, {Barsotti}, {Barsuglia},
  {Barta}, {Bartlett}, {Bartos}, {Bassiri}, {Basti}, {Bawaj}, {Bayley},
  {Bazzan}, {B{\'e}csy}, {Bejger}, {Belahcene}, {Bell}, {Beniwal}, {Berger},
  {Bergmann}, {Bernuzzi}, {Bero}, {Berry}, {Bersanetti}, {Bertolini},
  {Betzwieser}, {Bhand are}, {Bidler}, {Bilenko}, {Bilgili}, {Billingsley},
  {Birch}, {Birney}, {Birnholtz}, {Biscans}, {Biscoveanu}, {Bisht}, {Bitossi},
  {Bizouard}, {Blackburn}, {Blair}, {Blair}, {Blair}, {Bloemen}, {Bode},
  {Boer}, {Boetzel}, {Bogaert}, {Bondu}, {Bonilla}, {Bonnand}, {Booker},
  {Boom}, {Booth}, {Bork}, {Boschi}, {Bose}, {Bossie}, {Bossilkov}, {Bosveld},
  {Bouffanais}, {Bozzi}, {Bradaschia}, {Brady}, {Bramley}, {Branchesi}, {Brau},
  {Briant}, {Briggs}, {Brighenti}, {Brillet}, {Brinkmann}, {Brisson},
  {Brockill}, {Brooks}, {Brown}, {Brunett}, {Buikema}, {Bulik}, {Bulten},
  {Buonanno}, {Buskulic}, {Buy}, {Byer}, {Cabero}, {Cadonati}, {Cagnoli},
  {Cahillane}, {Calder{\'o}n Bustillo}, {Callister}, {Calloni}, {Camp},
  {Campbell}, {Cannon}, {Cao}, {Cao}, {Capocasa}, {Carbognani}, {Caride},
  {Carney}, {Carullo}, {Casanueva Diaz}, {Casentini}, {Caudill},
  {Cavagli{\`a}}, {Cavalier}, {Cavalieri}, {Cella}, {Cerd{\'a}-Dur{\'a}n},
  {Cerretani}, {Cesarini}, {Chaibi}, {Chakravarti}, {Chamberlin}, {Chan},
  {Chao}, {Charlton}, {Chase}, {Chassande-Mottin}, {Chatterjee}, {Chaturvedi},
  {Chatziioannou}, {Cheeseboro}, {Chen}, {Chen}, {Chen}, {Cheng}, {Cheong},
  {Chia}, {Chincarini}, {Chiummo}, {Cho}, {Cho}, {Cho}, {Christensen}, {Chu},
  {Chua}, {Chung}, {Chung}, {Ciani}, {Ciecielag}, {Ciobanu}, {Ciolfi},
  {Cipriano}, {Cirone}, {Clara}, {Clark}, {Clearwater}, {Cleva}, {Cocchieri},
  {Coccia}, {Cohadon}, {Cohen}, {Colgan}, {Colleoni}, {Collette}, {Collins},
  {Cominsky}, {Constancio}, {Conti}, {Cooper}, {Corban}, {Corbitt},
  {Cordero-Carri{\'o}n}, {Corley}, {Cornish}, {Corsi}, {Cortese}, {Costa},
  {Cotesta}, {Coughlin}, {Coughlin}, {Coulon}, {Countryman}, {Couvares},
  {Covas}, {Cowan}, {Coward}, {Cowart}, {Coyne}, {Coyne}, {Creighton},
  {Creighton}, {Cripe}, {Croquette}, {Crowder}, {Cullen}, {Cumming},
  {Cunningham}, {Cuoco}, {Dal Canton}, {D{\'a}lya}, {Danilishin}, {D'Antonio},
  {Danzmann}, {Dasgupta}, {Da Silva Costa}, {Datrier}, {Dattilo}, {Dave},
  {Davier}, {Davis}, {Daw}, {DeBra}, {Deenadayalan}, {Degallaix}, {De
  Laurentis}, {Del{\'e}glise}, {Del Pozzo}, {DeMarchi}, {Demos}, {Dent}, {De
  Pietri}, {Derby}, {De Rosa}, {De Rossi}, {DeSalvo}, {de Varona},
  {Dhurandhar}, {D{\'\i}az}, {Dietrich}, {Di Fiore}, {Di Giovanni}, {Di
  Girolamo}, {Di Lieto}, {Ding}, {Di Pace}, {Di Palma}, {Di Renzo}, {Dmitriev},
  {Doctor}, {Donovan}, {Dooley}, {Doravari}, {Dorosh}, {Dorrington}, {Downes},
  {Drago}, {Driggers}, {Du}, {Ducoin}, {Dupej}, {Dwyer}, {Easter}, {Edo},
  {Edwards}, {Effler}, {Ehrens}, {Eichholz}, {Eikenberry}, {Eisenmann},
  {Eisenstein}, {Essick}, {Estelles}, {Estevez}, {Etienne}, {Etzel}, {Evans},
  {Evans}, {Fafone}, {Fair}, {Fairhurst}, {Fan}, {Farinon}, {Farr}, {Farr},
  {Fauchon-Jones}, {Favata}, {Fays}, {Fazio}, {Fee}, {Feicht}, {Fejer}, {Feng},
  {Fernand ez-Galiana}, {Ferrante}, {Ferreira}, {Ferreira}, {Ferrini},
  {Fidecaro}, {Fiori}, {Fiorucci}, {Fishbach}, {Fisher}, {Fishner},
  {Fitz-Axen}, {Flaminio}, {Fletcher}, {Flynn}, {Fong}, {Font}, {Forsyth},
  {Fournier}, {Frasca}, {Frasconi}, {Frei}, {Freise}, {Frey}, {Frey},
  {Fritschel}, {Frolov}, {Fulda}, {Fyffe}, {Gabbard}, {Gadre}, {Gaebel},
  {Gair}, {Gammaitoni}, {Ganija}, {Gaonkar}, {Garcia},
  {Garc{\'\i}a-Quir{\'o}s}, {Garufi}, {Gateley}, {Gaudio}, {Gaur}, {Gayathri},
  {Gemme}, {Genin}, {Gennai}, {George}, {George}, {Gergely}, {Germain},
  {Ghonge}, {Ghosh}, {Ghosh}, {Ghosh}, {Giacomazzo}, {Giaime}, {Giardina},
  {Giazotto}, {Gill}, {Giordano}, {Glover}, {Godwin}, {Goetz}, {Goetz},
  {Goncharov}, {Gonz{\'a}lez}, {Gonzalez Castro}, {Gopakumar}, {Gorodetsky},
  {Gossan}, {Gosselin}, {Gouaty}, {Grado}, {Graef}, {Granata}, {Grant}, {Gras},
  {Grassia}, {Gray}, {Gray}, {Greco}, {Green}, {Green}, {Gretarsson}, {Groot},
  {Grote}, {Grunewald}, {Gruning}, {Guidi}, {Gulati}, {Guo}, {Gupta}, {Gupta},
  {Gustafson}, {Gustafson}, {Haegel}, {Halim}, {Hall}, {Hall}, {Hamilton},
  {Hammond}, {Haney}, {Hanke}, {Hanks}, {Hanna}, {Hannam}, {Hannuksela},
  {Hanson}, {Hardwick}, {Haris}, {Harms}, {Harry}, {Harry}, {Haskell},
  {Haster}, {Haughian}, {Hayes}, {Healy}, {Heidmann}, {Heintze}, {Heitmann},
  {Hello}, {Hemming}, {Hendry}, {Heng}, {Hennig}, {Heptonstall}, {Hernandez
  Vivanco}, {Heurs}, {Hild}, {Hinderer}, {Hoak}, {Hochheim}, {Hofman},
  {Holgado}, {Holland }, {Holt}, {Holz}, {Hopkins}, {Horst}, {Hough},
  {Hourihane}, {Howell}, {Hoy}, {Hreibi}, {Huerta}, {Huet}, {Hughey}, {Hulko},
  {Husa}, {Huttner}, {Huynh-Dinh}, {Idzkowski}, {Iess}, {Ingram}, {Inta},
  {Intini}, {Irwin}, {Isa}, {Isac}, {Isi}, {Iyer}, {Izumi}, {Jacqmin},
  {Jadhav}, {Jani}, {Janthalur}, {Jaranowski}, {Jenkins}, {Jiang}, {Johnson},
  {Jones}, {Jones}, {Jones}, {Jonker}, {Ju}, {Junker}, {Kalaghatgi},
  {Kalogera}, {Kamai}, {Kand hasamy}, {Kang}, {Kanner}, {Kapadia}, {Karki},
  {Karvinen}, {Kashyap}, {Kasprzack}, {Katsanevas}, {Katsavounidis}, {Katzman},
  {Kaufer}, {Kawabe}, {Keerthana}, {K{\'e}f{\'e}lian}, {Keitel}, {Kennedy},
  {Key}, {Khalili}, {Khan}, {Khan}, {Khan}, {Khan}, {Khazanov}, {Khursheed},
  {Kijbunchoo}, {Kim}, {Kim}, {Kim}, {Kim}, {Kim}, {Kim}, {Kimball}, {King},
  {King}, {Kinley-Hanlon}, {Kirchhoff}, {Kissel}, {Kleybolte}, {Klika},
  {Klimenko}, {Knowles}, {Koch}, {Koehlenbeck}, {Koekoek}, {Koley},
  {Kondrashov}, {Kontos}, {Koper}, {Korobko}, {Korth}, {Kowalska}, {Kozak},
  {Kringel}, {Krishnendu}, {Kr{\'o}lak}, {Kuehn}, {Kumar}, {Kumar}, {Kumar},
  {Kumar}, {Kuo}, {Kutynia}, {Kwang}, {Lackey}, {Lai}, {Lam}, {Landry}, {Lane},
  {Lang}, {Lange}, {Lantz}, {Lanza}, {Lartaux-Vollard}, {Lasky}, {Laxen},
  {Lazzarini}, {Lazzaro}, {Leaci}, {Leavey}, {Lecoeuche}, {Lee}, {Lee}, {Lee},
  {Lee}, {Lee}, {Lee}, {Lehmann}, {Lenon}, {Leroy}, {Letendre}, {Levin},
  {Leviton}, {Li}, {Li}, {Li}, {Li}, {Lin}, {Linde}, {Linker}, {Littenberg},
  {Liu}, {Liu}, {Lo}, {Lockerbie}, {London}, {Longo}, {Lorenzini}, {Loriette},
  {Lormand}, {Losurdo}, {Lough}, {Lousto}, {Lovelace}, {Lower}, {L{\"u}ck},
  {Lumaca}, {Lundgren}, {Lynch}, {Ma}, {Macas}, {Macfoy}, {MacInnis},
  {Macleod}, {Macquet}, {Maga{\~n}a-Sandoval}, {Maga{\~n}a Zertuche}, {Magee},
  {Majorana}, {Maksimovic}, {Malik}, {Man}, {Mandic}, {Mangano}, {Mansell},
  {Manske}, {Mantovani}, {Marchesoni}, {Marion}, {M{\'a}rka}, {M{\'a}rka},
  {Markakis}, {Markosyan}, {Markowitz}, {Maros}, {Marquina}, {Marsat},
  {Martelli}, {Martin}, {Martin}, {Martynov}, {Mason}, {Massera}, {Masserot},
  {Massinger}, {Masso-Reid}, {Mastrogiovanni}, {Matas}, {Matichard}, {Matone},
  {Mavalvala}, {Mazumder}, {McCann}, {McCarthy}, {McClelland }, {Zanolin},
  {Zelenova}, {Zendri}, {Zevin}, {Zhang}, {Zhang}, {Zhang}, {Zhao}, {Zhou},
  {Zhou}, {Zhu}, {Zucker}, {Zweizig}, {Pisarski}, {LIGO Scientific
  Collaboration}, \& {Virgo Collaboration}}]{all-sky:2019}
{Abbott}, B.~P., {Abbott}, R., {Abbott}, T.~D., {et~al.} 2019{\natexlab{a}},
  \href{http://dx.doi.org/10.1103/PhysRevD.100.024004}{\JournalTitle{\prd},
  100, 024004}

\bibitem[{{Abbott} {et~al.}(2019{\natexlab{b}}){Abbott}, {Abbott}, {Abbott},
  {Abraham}, {Acernese}, {Ackley}, {Adams}, {Adhikari}, {Adya}, {Affeldt},
  {Agathos}, {Agatsuma}, {Aggarwal}, {Aguiar}, {Aiello}, {Ain}, {Ajith},
  {Allen}, {Allocca}, {Aloy}, {Altin}, {Amato}, {Ananyeva}, {Anderson},
  {Anderson}, {Angelova}, {Antier}, {Appert}, {Arai}, {Araya}, {Areeda},
  {Ar{\`e}ne}, {Arnaud}, {Ascenzi}, {Ashton}, {Aston}, {Astone}, {Aubin},
  {Aufmuth}, {AultONeal}, {Austin}, {Avendano}, {Avila-Alvarez}, {Babak},
  {Bacon}, {Badaracco}, {Bader}, {Bae}, {Bailes}, {Baker}, {Baldaccini},
  {Ballardin}, {Ballmer}, {Banagiri}, {Barayoga}, {Barclay}, {Barish},
  {Barker}, {Barkett}, {Barnum}, {Barone}, {Barr}, {Barsotti}, {Barsuglia},
  {Barta}, {Bartlett}, {Bartos}, {Bassiri}, {Basti}, {Bawaj}, {Bayley},
  {Bazzan}, {B{\'e}csy}, {Bejger}, {Belahcene}, {Bell}, {Beniwal}, {Berger},
  {Bergmann}, {Bernuzzi}, {Bero}, {Berry}, {Bersanetti}, {Bertolini},
  {Betzwieser}, {Bhandare}, {Bidler}, {Bilenko}, {Bilgili}, {Billingsley},
  {Birch}, {Birney}, {Birnholtz}, {Biscans}, {Biscoveanu}, {Bisht}, {Bitossi},
  {Bizouard}, {Blackburn}, {Blair}, {Blair}, {Blair}, {Bloemen}, {Bode},
  {Boer}, {Boetzel}, {Bogaert}, {Bondu}, {Bonilla}, {Bonnand}, {Booker},
  {Boom}, {Booth}, {Bork}, {Boschi}, {Bose}, {Bossie}, {Bossilkov}, {Bosveld},
  {Bouffanais}, {Bozzi}, {Bradaschia}, {Brady}, {Bramley}, {Branchesi}, {Brau},
  {Briant}, {Briggs}, {Brighenti}, {Brillet}, {Brinkmann}, {Brisson},
  {Brockill}, {Brooks}, {Brown}, {Brunett}, {Buikema}, {Bulik}, {Bulten},
  {Buonanno}, {Buskulic}, {Buy}, {Byer}, {Cabero}, {Cadonati}, {Cagnoli},
  {Cahillane}, {Calder{\'o}n Bustillo}, {Callister}, {Calloni}, {Camp},
  {Campbell}, {Canepa}, {Cannon}, {Cao}, {Cao}, {Capocasa}, {Carbognani},
  {Caride}, {Carney}, {Carullo}, {Casanueva Diaz}, {Casentini}, {Caudill},
  {Cavagli{\`a}}, {Cavalier}, {Cavalieri}, {Cella}, {Cerd{\'a}-Dur{\'a}n},
  {Cerretani}, {Cesarini}, {Chaibi}, {Chakravarti}, {Chamberlin}, {Chan},
  {Chao}, {Charlton}, {Chase}, {Chassande-Mottin}, {Chatterjee}, {Chaturvedi},
  {Cheeseboro}, {Chen}, {Chen}, {Chen}, {Cheng}, {Cheong}, {Chia},
  {Chincarini}, {Chiummo}, {Cho}, {Cho}, {Cho}, {Christensen}, {Chu}, {Chua},
  {Chung}, {Chung}, {Ciani}, {Ciobanu}, {Ciolfi}, {Cipriano}, {Cirone},
  {Clara}, {Clark}, {Clearwater}, {Cleva}, {Cocchieri}, {Coccia}, {Cohadon},
  {Cohen}, {Colgan}, {Colleoni}, {Collette}, {Collins}, {Cominsky},
  {Constancio}, {Conti}, {Cooper}, {Corban}, {Corbitt}, {Cordero-Carri{\'o}n},
  {Corley}, {Cornish}, {Corsi}, {Cortese}, {Costa}, {Cotesta}, {Coughlin},
  {Coughlin}, {Coulon}, {Countryman}, {Couvares}, {Covas}, {Cowan}, {Coward},
  {Cowart}, {Coyne}, {Coyne}, {Creighton}, {Creighton}, {Cripe}, {Croquette},
  {Crowder}, {Cullen}, {Cumming}, {Cunningham}, {Cuoco}, {Dal Canton},
  {D{\'a}lya}, {Danilishin}, {D'Antonio}, {Danzmann}, {Dasgupta}, {Da Silva
  Costa}, {Datrier}, {Dattilo}, {Dave}, {Davier}, {Davis}, {Daw}, {DeBra},
  {Deenadayalan}, {Degallaix}, {De Laurentis}, {Del{\'e}glise}, {Del Pozzo},
  {DeMarchi}, {Demos}, {Dent}, {De Pietri}, {Derby}, {De Rosa}, {De Rossi},
  {DeSalvo}, {de Varona}, {Dhurandhar}, {D{\'\i}az}, {Dietrich}, {Di Fiore},
  {Di Giovanni}, {Di Girolamo}, {Di Lieto}, {Ding}, {Di Pace}, {Di Palma}, {Di
  Renzo}, {Dmitriev}, {Doctor}, {Donovan}, {Dooley}, {Doravari}, {Dorrington},
  {Downes}, {Drago}, {Driggers}, {Du}, {Ducoin}, {Dupej}, {Dwyer}, {Easter},
  {Edo}, {Edwards}, {Effler}, {Ehrens}, {Eichholz}, {Eikenberry}, {Eisenmann},
  {Eisenstein}, {Essick}, {Estelles}, {Estevez}, {Etienne}, {Etzel}, {Evans},
  {Evans}, {Fafone}, {Fair}, {Fairhurst}, {Fan}, {Farinon}, {Farr}, {Farr},
  {Fauchon-Jones}, {Favata}, {Fays}, {Fazio}, {Fee}, {Feicht}, {Fejer}, {Feng},
  {Fernandez-Galiana}, {Ferrante}, {Ferreira}, {Ferreira}, {Ferrini},
  {Fidecaro}, {Fiori}, {Fiorucci}, {Fishbach}, {Fisher}, {Fishner},
  {Fitz-Axen}, {Flaminio}, {Fletcher}, {Flynn}, {Fong}, {Font}, {Forsyth},
  {Fournier}, {Frasca}, {Frasconi}, {Frei}, {Freise}, {Frey}, {Frey},
  {Fritschel}, {Frolov}, {Fulda}, {Fyffe}, {Gabbard}, {Gadre}, {Gaebel},
  {Gair}, {Gammaitoni}, {Ganija}, {Gaonkar}, {Garcia},
  {Garc{\'\i}a-Quir{\'o}s}, {Garufi}, {Gateley}, {Gaudio}, {Gaur}, {Gayathri},
  {Gemme}, {Genin}, {Gennai}, {George}, {George}, {Gergely}, {Germain},
  {Ghonge}, {Ghosh}, {Ghosh}, {Ghosh}, {Giacomazzo}, {Giaime}, {Giardina},
  {Giazotto}, {Gill}, {Giordano}, {Glover}, {Godwin}, {Goetz}, {Goetz},
  {Goncharov}, {Gonz{\'a}lez}, {Gonzalez Castro}, {Gopakumar}, {Gorodetsky},
  {Gossan}, {Gosselin}, {Gouaty}, {Grado}, {Graef}, {Granata}, {Grant}, {Gras},
  {Grassia}, {Gray}, {Gray}, {Greco}, {Green}, {Green}, {Gretarsson}, {Groot},
  {Grote}, {Grunewald}, {Gruning}, {Guidi}, {Gulati}, {Guo}, {Gupta}, {Gupta},
  {Gustafson}, {Gustafson}, {Haegel}, {Halim}, {Hall}, {Hall}, {Hamilton},
  {Hammond}, {Haney}, {Hanke}, {Hanks}, {Hanna}, {Hannam}, {Hannuksela},
  {Hanson}, {Hardwick}, {Haris}, {Harms}, {Harry}, {Harry}, {Haster},
  {Haughian}, {Hayes}, {Healy}, {Heidmann}, {Heintze}, {Heitmann}, {Hello},
  {Hemming}, {Hendry}, {Heng}, {Hennig}, {Heptonstall}, {Hernandez Vivanco},
  {Heurs}, {Hild}, {Hinderer}, {Ho}, {Hoak}, {Hochheim}, {Hofman}, {Holgado},
  {Holland}, {Holt}, {Holz}, {Hopkins}, {Horst}, {Hough}, {Howell}, {Hoy},
  {Hreibi}, {Huerta}, {Huet}, {Hughey}, {Hulko}, {Husa}, {Huttner},
  {Huynh-Dinh}, {Idzkowski}, {Iess}, {Ingram}, {Inta}, {Intini}, {Irwin},
  {Isa}, {Isac}, {Isi}, {Iyer}, {Izumi}, {Jacqmin}, {Jadhav}, {Jani},
  {Janthalur}, {Jaranowski}, {Jenkins}, {Jiang}, {Johnson}, {Jones}, {Jones},
  {Jones}, {Jonker}, {Ju}, {Junker}, {Kalaghatgi}, {Kalogera}, {Kamai},
  {Kandhasamy}, {Kang}, {Kanner}, {Kapadia}, {Karki}, {Karvinen}, {Kashyap},
  {Kasprzack}, {Katsanevas}, {Katsavounidis}, {Katzman}, {Kaufer}, {Kawabe},
  {Keerthana}, {K{\'e}f{\'e}lian}, {Keitel}, {Kennedy}, {Key}, {Khalili},
  {Khan}, {Khan}, {Khan}, {Khan}, {Khazanov}, {Khursheed}, {Kijbunchoo}, {Kim},
  {Kim}, {Kim}, {Kim}, {Kim}, {Kim}, {Kimball}, {King}, {King},
  {Kinley-Hanlon}, {Kirchhoff}, {Kissel}, {Kleybolte}, {Klika}, {Klimenko},
  {Knowles}, {Koch}, {Koehlenbeck}, {Koekoek}, {Koley}, {Kondrashov}, {Kontos},
  {Koper}, {Korobko}, {Korth}, {Kowalska}, {Kozak}, {Kringel}, {Krishnendu},
  {Kr{\'o}lak}, {Kuehn}, {Kumar}, {Kumar}, {Kumar}, {Kumar}, {Kuo}, {Kutynia},
  {Kwang}, {Lackey}, {Lai}, {Lam}, {Landry}, {Lane}, {Lang}, {Lange}, {Lantz},
  {Lanza}, {Lartaux-Vollard}, {Lasky}, {Laxen}, {Lazzarini}, {Lazzaro},
  {Leaci}, {Leavey}, {Lecoeuche}, {Lee}, {Lee}, {Lee}, {Lee}, {Lee}, {Lee},
  {Lehmann}, {Lenon}, {Leroy}, {Letendre}, {Levin}, {Li}, {Li}, {Li}, {Li},
  {Lin}, {Linde}, {Linker}, {Littenberg}, {Liu}, {Liu}, {Lo}, {Lockerbie},
  {London}, {Longo}, {Lorenzini}, {Loriette}, {Lormand}, {Losurdo}, {Lough},
  {Lousto}, {Lovelace}, {Lower}, {L{\"u}ck}, {Lumaca}, {Lundgren}, {Lynch},
  {Ma}, {Macas}, {Macfoy}, {MacInnis}, {Macleod}, {Macquet},
  {Maga{\~n}a-Sandoval}, {Maga{\~n}a Zertuche}, {Magee}, {Majorana},
  {Maksimovic}, {Malik}, {Man}, {Mandic}, {Mangano}, {Mansell}, {Manske},
  {Mantovani}, {Marchesoni}, {Marion}, {M{\'a}rka}, {M{\'a}rka}, {Markakis},
  {Markosyan}, {Markowitz}, {Maros}, {Marquina}, {Marsat}, {Martelli},
  {Martin}, {Martin}, {Martynov}, {Mason}, {Massera}, {Masserot}, {Massinger},
  {Masso-Reid}, {Mastrogiovanni}, {Matas}, {Matichard}, {Matone}, {Mavalvala},
  {Mazumder}, {McCann}, {McCarthy}, {McClelland}, {McCormick}, {McCuller},
  {McGuire}, {McIver}, {McManus}, {McRae}, {McWilliams}, {Meacher}, {Meadors},
  {Mehmet}, {Mehta}, {Meidam}, {Melatos}, {Mendell}, {Mercer}, {Mereni},
  {Merilh}, {Merzougui}, {Meshkov}, {Messenger}, {Messick}, {Metzdorff},
  {Meyers}, {Miao}, {Michel}, {Middleton}, {Mikhailov}, {Milano}, {Miller},
  {Miller}, {Millhouse}, {Mills}, {Milovich-Goff}, {Minazzoli}, {Minenkov},
  {Mishkin}, {Mishra}, {Mistry}, {Mitra}, {Mitrofanov}, {Mitselmakher},
  {Mittleman}, {Mo}, {Moffa}, {Mogushi}, {Mohapatra}, {Montani}, {Moore},
  {Moraru}, {Moreno}, {Morisaki}, {Mours}, {Mow-Lowry}, {Mukherjee},
  {Mukherjee}, {Mukherjee}, {Mukund}, {Mullavey}, {Munch}, {Mu{\~n}iz},
  {Muratore}, {Murray}, {Nagar}, {Nardecchia}, {Naticchioni}, {Nayak},
  {Neilson}, {Nelemans}, {Nelson}, {Nery}, {Neunzert}, {Ng}, {Ng}, {Nguyen},
  {Nichols}, {Nissanke}, {Nocera}, {North}, {Nuttall}, {Obergaulinger},
  {Oberling}, {O'Brien}, {O'Dea}, {Ogin}, {Oh}, {Oh}, {Ohme}, {Ohta}, {Okada},
  {Oliver}, {Oppermann}, {Oram}, {O'Reilly}, {Ormiston}, {Ortega},
  {O'Shaughnessy}, {Ossokine}, {Ottaway}, {Overmier}, {Owen}, {Pace}, {Pagano},
  {Page}, {Pai}, {Pai}, {Palamos}, {Palashov}, {Palomba}, {Pal-Singh}, {Pan},
  {Pang}, {Pang}, {Pankow}, {Pannarale}, {Pant}, {Paoletti}, {Paoli}, {Parida},
  {Parker}, {Pascucci}, {Pasqualetti}, {Passaquieti}, {Passuello}, {Patil},
  {Patricelli}, {Pearlstone}, {Pedersen}, {Pedraza}, {Pedurand}, {Pele},
  {Penn}, {Perez}, {Perreca}, {Pfeiffer}, {Phelps}, {Phukon}, {Piccinni},
  {Pichot}, {Piergiovanni}, {Pillant}, {Pinard}, {Pirello}, {Pitkin},
  {Poggiani}, {Pong}, {Ponrathnam}, {Popolizio}, {Porter}, {Powell},
  {Prajapati}, {Prasad}, {Prasai}, {Prasanna}, {Pratten}, {Prestegard},
  {Privitera}, {Prodi}, {Prokhorov}, {Puncken}, {Punturo}, {Puppo},
  {P{\"u}rrer}, {Qi}, {Quetschke}, {Quinonez}, {Quintero}, {Quitzow-James},
  {Raab}, {Radkins}, {Radulescu}, {Raffai}, {Raja}, {Rajan}, {Rajbhandari},
  {Rakhmanov}, {Ramirez}, {Ramos-Buades}, {Rana}, {Rao}, {Rapagnani},
  {Raymond}, {Razzano}, {Read}, {Regimbau}, {Rei}, {Reid}, {Reitze}, {Ren},
  {Ricci}, {Richardson}, {Richardson}, {Ricker}, {Riles}, {Rizzo}, {Robertson},
  {Robie}, {Robinet}, {Rocchi}, {Rolland}, {Rollins}, {Roma}, {Romanelli},
  {Romano}, {Romel}, {Romie}, {Rose}, {Rosi{\'n}ska}, {Rosofsky}, {Ross},
  {Rowan}, {R{\"u}diger}, {Ruggi}, {Rutins}, {Ryan}, {Sachdev}, {Sadecki},
  {Sakellariadou}, {Salconi}, {Saleem}, {Samajdar}, {Sammut}, {Sanchez},
  {Sanchez}, {Sanchis-Gual}, {Sandberg}, {Sanders}, {Santiago}, {Sarin},
  {Sassolas}, {Saulson}, {Sauter}, {Savage}, {Schale}, {Scheel}, {Scheuer},
  {Schmidt}, {Schnabel}, {Schofield}, {Sch{\"o}nbeck}, {Schreiber}, {Schulte},
  {Schutz}, {Schwalbe}, {Scott}, {Scott}, {Seidel}, {Sellers}, {Sengupta},
  {Sennett}, {Sentenac}, {Sequino}, {Sergeev}, {Setyawati}, {Shaddock},
  {Shaffer}, {Shahriar}, {Shaner}, {Shao}, {Sharma}, {Shawhan}, {Shen},
  {Shink}, {Shoemaker}, {Shoemaker}, {ShyamSundar}, {Siellez}, {Sieniawska},
  {Sigg}, {Silva}, {Singer}, {Singh}, {Singhal}, {Sintes}, {Sitmukhambetov},
  {Skliris}, {Slagmolen}, {Slaven-Blair}, {Smith}, {Smith}, {Somala}, {Son},
  {Sorazu}, {Sorrentino}, {Souradeep}, {Sowell}, {Spencer}, {Srivastava},
  {Srivastava}, {Staats}, {Stachie}, {Standke}, {Steer}, {Steinke},
  {Steinlechner}, {Steinlechner}, {Steinmeyer}, {Stevenson}, {Stocks}, {Stone},
  {Stops}, {Strain}, {Stratta}, {Strigin}, {Strunk}, {Sturani}, {Stuver},
  {Sudhir}, {Summerscales}, {Sun}, {Sunil}, {Suresh}, {Sutton}, {Swinkels},
  {Szczepa{\'n}czyk}, {Tacca}, {Tait}, {Talbot}, {Talukder}, {Tanner},
  {T{\'a}pai}, {Taracchini}, {Tasson}, {Taylor}, {Thies}, {Thomas}, {Thomas},
  {Thondapu}, {Thorne}, {Thrane}, {Tiwari}, {Tiwari}, {Tiwari}, {Toland},
  {Tonelli}, {Tornasi}, {Torres-Forn{\'e}}, {Torrie}, {T{\"o}yr{\"a}},
  {Travasso}, {Traylor}, {Tringali}, {Trovato}, {Trozzo}, {Trudeau}, {Tsang},
  {Tse}, {Tso}, {Tsukada}, {Tsuna}, {Tuyenbayev}, {Ueno}, {Ugolini},
  {Unnikrishnan}, {Urban}, {Usman}, {Vahlbruch}, {Vajente}, {Valdes}, {van
  Bakel}, {van Beuzekom}, {van den Brand}, {Van Den Broeck}, {Vander-Hyde},
  {van Heijningen}, {van der Schaaf}, {van Veggel}, {Vardaro}, {Varma}, {Vass},
  {Vas{\'u}th}, {Vecchio}, {Vedovato}, {Veitch}, {Veitch}, {Venkateswara},
  {Venugopalan}, {Verkindt}, {Vetrano}, {Vicer{\'e}}, {Viets}, {Vine}, {Vinet},
  {Vitale}, {Vo}, {Vocca}, {Vorvick}, {Vyatchanin}, {Wade}, {Wade}, {Wade},
  {Walet}, {Walker}, {Wallace}, {Walsh}, {Wang}, {Wang}, {Wang}, {Wang},
  {Wang}, {Ward}, {Warden}, {Warner}, {Was}, {Watchi}, {Weaver}, {Wei},
  {Weinert}, {Weinstein}, {Weiss}, {Wellmann}, {Wen}, {Wessel}, {We{\ss}els},
  {Westhouse}, {Wette}, {Whelan}, {Whiting}, {Whittle}, {Wilken}, {Williams},
  {Williamson}, {Willis}, {Willke}, {Wimmer}, {Winkler}, {Wipf}, {Wittel},
  {Woan}, {Woehler}, {Wofford}, {Worden}, {Wright}, {Wu}, {Wysocki}, {Xiao},
  {Yamamoto}, {Yancey}, {Yang}, {Yap}, {Yazback}, {Yeeles}, {Yu}, {Yu}, {Yuen},
  {Yvert}, {Zadro{\.z}ny}, {Zanolin}, {Zelenova}, {Zendri}, {Zevin}, {Zhang},
  {Zhang}, {Zhang}, {Zhao}, {Zhou}, {Zhou}, {Zhu}, {Zucker}, {Zweizig}, {LIGO
  Scientific Collaboration}, {Virgo Collaboration}, {Arzoumanian}, {Bogdanov},
  {Cognard}, {Corongiu}, {Enoto}, {Freire}, {Gendreau}, {Guillemot}, {Harding},
  {Jankowski}, {Keith}, {Kerr}, {Lyne}, {Palfreyman}, {Possenti}, {Ridolfi},
  {Stappers}, {Theureau}, \& {Weltervrede}}]{spin-down}
---. 2019{\natexlab{b}},
  \href{http://dx.doi.org/10.3847/1538-4357/ab20cb}{\JournalTitle{\apj}, 879,
  10}

\bibitem[{Abbott {et~al.}(2019b)Abbott, Abbott, Abbott, Abraham, Acernese,
  Ackley, Adams, Adhikari, Adya, Affeldt, Agathos, Agatsuma, Aggarwal, Aguiar,
  Aiello, Ain, Ajith, Allen, Allocca, Aloy, Altin, Amato, Ananyeva, Anderson,
  Anderson, Angelova, Antier, Appert, Arai, Araya, Areeda, Ar\`ene, Arnaud,
  Ascenzi, Ashton, Aston, Astone, Aubin, Aufmuth, AultONeal, Austin, Avendano,
  Avila-Alvarez, Babak, Bacon, Badaracco, Bader, Bae, Baker, Baldaccini,
  Ballardin, Ballmer, Banagiri, Barayoga, Barclay, Barish, Barker, Barkett,
  Barnum, Barone, Barr, Barsotti, Barsuglia, Barta, Bartlett, Bartos, Bassiri,
  Basti, Bawaj, Bayley, Bazzan, B\'ecsy, Bejger, Belahcene, Bell, Beniwal,
  Berger, Bergmann, Bernuzzi, Bero, Berry, Bersanetti, Bertolini, Betzwieser,
  Bhandare, Bidler, Bilenko, Bilgili, Billingsley, Birch, Birney, Birnholtz,
  Biscans, Biscoveanu, Bisht, Bitossi, Bizouard, Blackburn, Blair, Blair,
  Blair, Bloemen, Bode, Boer, Boetzel, Bogaert, Bondu, Bonilla, Bonnand,
  Booker, Boom, Booth, Bork, Boschi, Bose, Bossie, Bossilkov, Bosveld,
  Bouffanais, Bozzi, Bradaschia, Brady, Bramley, Branchesi, Brau, Briant,
  Briggs, Brighenti, Brillet, Brinkmann, Brisson, Brockill, Brooks, Brown,
  Brunett, Buikema, Bulik, Bulten, Buonanno, Buskulic, Buy, Byer, Cabero,
  Cadonati, Cagnoli, Cahillane, Bustillo, Callister, Calloni, Camp, Campbell,
  Canepa, Cannon, Cao, Cao, Capocasa, Carbognani, Caride, Carney, Carullo,
  Diaz, Casentini, Caudill, Cavagli\`a, Cavalier, Cavalieri, Cella,
  Cerd\'a-Dur\'an, Cerretani, Cesarini, Chaibi, Chakravarti, Chamberlin, Chan,
  Chao, Charlton, Chase, Chassande-Mottin, Chatterjee, Chaturvedi, Cheeseboro,
  Chen, Chen, Chen, Cheng, Cheong, Chia, Chincarini, Chiummo, Cho, Cho, Cho,
  Christensen, Chu, Chua, Chung, Chung, Ciani, Ciobanu, Ciolfi, Cipriano,
  Cirone, Clara, Clark, Clearwater, Cleva, Cocchieri, Coccia, Cohadon, Cohen,
  Colgan, Colleoni, Collette, Collins, Cominsky, Constancio, Conti, Cooper,
  Corban, Corbitt, Cordero-Carri\'on, Corley, Cornish, Corsi, Cortese, Costa,
  Cotesta, Coughlin, Coughlin, Coulon, Countryman, Couvares, Covas, Cowan,
  Coward, Cowart, Coyne, Coyne, Creighton, Creighton, Cripe, Croquette,
  Crowder, Cullen, Cumming, Cunningham, Cuoco, Canton, D\'alya, Danilishin,
  D'Antonio, Danzmann, Dasgupta, Costa, Datrier, Dattilo, Dave, Davier, Davis,
  Daw, DeBra, Deenadayalan, Degallaix, De~Laurentis, Del\'eglise, Del~Pozzo,
  DeMarchi, Demos, Dent, De~Pietri, Derby, De~Rosa, De~Rossi, DeSalvo,
  de~Varona, Dhurandhar, D\'{\i}az, Dietrich, Di~Fiore, Di~Giovanni,
  Di~Girolamo, Di~Lieto, Ding, Di~Pace, Di~Palma, Di~Renzo, Dmitriev, Doctor,
  Donovan, Dooley, Doravari, Dorrington, Downes, Drago, Driggers, Du, Ducoin,
  Dupej, Dwyer, Easter, Edo, Edwards, Effler, Ehrens, Eichholz, Eikenberry,
  Eisenmann, Eisenstein, Essick, Estelles, Estevez, Etienne, Etzel, Evans,
  Evans, Fafone, Fair, Fairhurst, Fan, Farinon, Farr, Farr, Fauchon-Jones,
  Favata, Fays, Fazio, Fee, Feicht, Fejer, Feng, Fernandez-Galiana, Ferrante,
  Ferreira, Ferreira, Ferrini, Fidecaro, Fiori, Fiorucci, Fishbach, Fisher,
  Fishner, Fitz-Axen, Flaminio, Fletcher, Flynn, Fong, Font, Forsyth, Fournier,
  Frasca, Frasconi, Frei, Freise, Frey, Frey, Fritschel, Frolov, Fulda, Fyffe,
  Gabbard, Gadre, Gaebel, Gair, Gammaitoni, Ganija, Gaonkar, Garcia,
  Garc\'{\i}a-Quir\'os, Garufi, Gateley, Gaudio, Gaur, Gayathri, Gemme, Genin,
  Gennai, George, George, Gergely, Germain, Ghonge, Ghosh, Ghosh, Ghosh,
  Giacomazzo, Giaime, Giardina, Giazotto, Gill, Giordano, Glover, Godwin,
  Goetz, Goetz, Goncharov, Gonz\'alez, Castro, Gopakumar, Gorodetsky, Gossan,
  Gosselin, Gouaty, Grado, Graef, Granata, Grant, Gras, Grassia, Gray, Gray,
  Greco, Green, Green, Gretarsson, Groot, Grote, Grunewald, Gruning, Guidi,
  Gulati, Guo, Gupta, Gupta, Gustafson, Gustafson, Haegel, Halim, Hall, Hall,
  Hamilton, Hammond, Haney, Hanke, Hanks, Hanna, Hannuksela, Hanson, Hardwick,
  Haris, Harms, Harry, Harry, Haster, Haughian, Hayes, Healy, Heidmann,
  Heintze, Heitmann, Hello, Hemming, Hendry, Heng, Hennig, Heptonstall,
  Vivanco, Heurs, Hild, Hinderer, Hoak, Hochheim, Hofman, Holgado, Holland,
  Holt, Holz, Hopkins, Horst, Hough, Howell, Hoy, Hreibi, Huerta, Huet, Hughey,
  Hulko, Husa, Huttner, Huynh-Dinh, Idzkowski, Iess, Ingram, Inta, Intini,
  Irwin, Isa, Isac, Isi, Iyer, Izumi, Jacqmin, Jadhav, Jani, Janthalur,
  Jaranowski, Jenkins, Jiang, Johnson, Jones, Jones, Jones, Jonker, Ju, Junker,
  Kalaghatgi, Kalogera, Kamai, Kandhasamy, Kang, Kanner, Kapadia, Karki,
  Karvinen, Kashyap, Kasprzack, Katsanevas, Katsavounidis, Katzman, Kaufer,
  Kawabe, Keerthana, K\'ef\'elian, Keitel, Kennedy, Key, Khalili, Khan, Khan,
  Khan, Khan, Khazanov, Khursheed, Kijbunchoo, Kim, Kim, Kim, Kim, Kim, Kim,
  Kimball, King, King, Kinley-Hanlon, Kirchhoff, Kissel, Kleybolte, Klika,
  Klimenko, Knowles, Koch, Koehlenbeck, Koekoek, Koley, Kondrashov, Kontos,
  Koper, Korobko, Korth, Kowalska, Kozak, Kringel, Krishnendu, Kr\'olak, Kuehn,
  Kumar, Kumar, Kumar, Kumar, Kuo, Kutynia, Kwang, Lackey, Lai, Lam, Landry,
  Lane, Lang, Lange, Lantz, Lanza, Lartaux-Vollard, Lasky, Laxen, Lazzarini,
  Lazzaro, Leaci, Leavey, Lecoeuche, Lee, Lee, Lee, Lee, Lee, Lee, Lehmann,
  Lenon, Leroy, Letendre, Levin, Li, Li, Li, Li, Lin, Linde, Linker,
  Littenberg, Liu, Liu, Lo, Lockerbie, London, Longo, Lorenzini, Loriette,
  Lormand, Losurdo, Lough, Lovelace, Lower, L\"uck, Lumaca, Lundgren, Lynch,
  Ma, Macas, Macfoy, MacInnis, Macleod, Macquet, Maga\~na Sandoval, Zertuche,
  Magee, Majorana, Maksimovic, Malik, Man, Mandic, Mangano, Mansell, Manske,
  Mantovani, Marchesoni, Marion, M\'arka, M\'arka, Markakis, Markosyan,
  Markowitz, Maros, Marquina, Marsat, Martelli, Martin, Martin, Martynov,
  Mason, Massera, Masserot, Massinger, Masso-Reid, Mastrogiovanni, Matas,
  Matichard, Matone, Mavalvala, Mazumder, McCann, McCarthy, McClelland,
  McCormick, McCuller, McGuire, McIver, McManus, McRae, McWilliams, Meacher,
  Meadors, Mehmet, Mehta, Meidam, Melatos, Mendell, Mercer, Mereni, Merilh,
  Merzougui, Meshkov, Messenger, Messick, Metzdorff, Meyers, Miao, Michel,
  Middleton, Mikhailov, Milano, Miller, Miller, Millhouse, Mills,
  Milovich-Goff, Minazzoli, Minenkov, Mishkin, Mishra, Mistry, Mitra,
  Mitrofanov, Mitselmakher, Mittleman, Mo, Moffa, Mogushi, Mohapatra, Montani,
  Moore, Moraru, Moreno, Morisaki, Mours, Mow-Lowry, Mukherjee, Mukherjee,
  Mukherjee, Mukund, Mullavey, Munch, Mu\~niz, Muratore, Murray, Nardecchia,
  Naticchioni, Nayak, Neilson, Nelemans, Nelson, Nery, Neunzert, Ng, Ng,
  Nguyen, Nichols, Nissanke, Nocera, North, Nuttall, Obergaulinger, Oberling,
  O'Brien, O'Dea, Ogin, Oh, Oh, Ohme, Ohta, Okada, Oliver, Oppermann, Oram,
  O'Reilly, Ormiston, Ortega, O'Shaughnessy, Ossokine, Ottaway, Overmier, Owen,
  Pace, Pagano, Page, Pai, Pai, Palamos, Palashov, Palomba, Pal-Singh, Pan,
  Pang, Pang, Pankow, Pannarale, Pant, Paoletti, Paoli, Parida, Parker,
  Pascucci, Pasqualetti, Passaquieti, Passuello, Patil, Patricelli, Pearlstone,
  Pedersen, Pedraza, Pedurand, Pele, Penn, Perez, Perreca, Pfeiffer, Phelps,
  Phukon, Piccinni, Pichot, Piergiovanni, Pillant, Pinard, Pirello, Pitkin,
  Poggiani, Pong, Ponrathnam, Popolizio, Porter, Powell, Prajapati, Prasad,
  Prasai, Prasanna, Pratten, Prestegard, Privitera, Prodi, Prokhorov, Puncken,
  Punturo, Puppo, P\"urrer, Qi, Quetschke, Quinonez, Quintero, Quitzow-James,
  Raab, Radkins, Radulescu, Raffai, Raja, Rajan, Rajbhandari, Rakhmanov,
  Ramirez, Ramos-Buades, Rana, Rao, Rapagnani, Raymond, Razzano, Read,
  Regimbau, Rei, Reid, Reitze, Ren, Ricci, Richardson, Richardson, Ricker,
  Riles, Rizzo, Robertson, Robie, Robinet, Rocchi, Rolland, Rollins, Roma,
  Romanelli, Romano, Romel, Romie, Rose, Rosi\ifmmode~\acute{n}\else
  \'{n}\fi{}ska, Rosofsky, Ross, Rowan, R\"udiger, Ruggi, Rutins, Ryan,
  Sachdev, Sadecki, Sakellariadou, Salconi, Saleem, Samajdar, Sammut, Sanchez,
  Sanchez, Sanchis-Gual, Sandberg, Sanders, Santiago, Sarin, Sassolas, Saulson,
  Sauter, Savage, Schale, Scheel, Scheuer, Schmidt, Schnabel, Schofield,
  Sch\"onbeck, Schreiber, Schulte, Schutz, Schwalbe, Scott, Scott, Seidel,
  Sellers, Sengupta, Sennett, Sentenac, Sequino, Sergeev, Setyawati, Shaddock,
  Shaffer, Shahriar, Shaner, Shao, Sharma, Shawhan, Shen, Shink, Shoemaker,
  Shoemaker, ShyamSundar, Siellez, Sieniawska, Sigg, Silva, Singer, Singh,
  Singhal, Sintes, Sitmukhambetov, Skliris, Slagmolen, Slaven-Blair, Smith,
  Smith, Somala, Son, Sorazu, Sorrentino, Souradeep, Sowell, Spencer,
  Srivastava, Srivastava, Staats, Stachie, Standke, Steer, Steinke,
  Steinlechner, Steinlechner, Steinmeyer, Stevenson, Stocks, Stone, Stops,
  Strain, Stratta, Strigin, Strunk, Sturani, Stuver, Sudhir, Summerscales, Sun,
  Sunil, Suresh, Sutton, Swinkels, Szczepa\ifmmode~\acute{n}\else
  \'{n}\fi{}czyk, Tacca, Tait, Talbot, Talukder, Tanner, T\'apai, Taracchini,
  Tasson, Taylor, Thies, Thomas, Thomas, Thondapu, Thorne, Thrane, Tiwari,
  Tiwari, Tiwari, Toland, Tonelli, Tornasi, Torres-Forn\'e, Torrie, T\"oyr\"a,
  Travasso, Traylor, Tringali, Trovato, Trozzo, Trudeau, Tsang, Tse, Tso,
  Tsukada, Tsuna, Tuyenbayev, Ueno, Ugolini, Unnikrishnan, Urban, Usman,
  Vahlbruch, Vajente, Valdes, van Bakel, van Beuzekom, van~den Brand, Van
  Den~Broeck, Vander-Hyde, van Heijningen, van~der Schaaf, van Veggel, Vardaro,
  Varma, Vass, Vas\'uth, Vecchio, Vedovato, Veitch, Veitch, Venkateswara,
  Venugopalan, Verkindt, Vetrano, Vicer\'e, Viets, Vine, Vinet, Vitale, Vo,
  Vocca, Vorvick, Vyatchanin, Wade, Wade, Wade, Walet, Walker, Wallace, Walsh,
  Wang, Wang, Wang, Wang, Wang, Ward, Warden, Warner, Was, Watchi, Weaver, Wei,
  Weinert, Weinstein, Weiss, Wellmann, Wen, Wessel, We\ss{}els, Westhouse,
  Wette, Whelan, Whiting, Whittle, Wilken, Williams, Williamson, Willis,
  Willke, Wimmer, Winkler, Wipf, Wittel, Woan, Woehler, Wofford, Worden,
  Wright, Wu, Wysocki, Xiao, Yamamoto, Yancey, Yang, Yap, Yazback, Yeeles, Yu,
  Yu, Yuen, Yvert, Zadro\ifmmode~\dot{z}\else \.{z}\fi{}ny, Zanolin, Zelenova,
  Zendri, Zevin, Zhang, Zhang, Zhang, Zhao, Zhou, Zhou, Zhu, Zucker, Zweizig,
  Keith, Kerr, Kuiper, Harding, Lyne, Palfreyman, Stappers, \&
  Weltervrede}]{PhysRevD.99.122002}
Abbott, B.~P., Abbott, R., Abbott, T.~D., {et~al.} 2019b,
  \href{http://dx.doi.org/10.1103/PhysRevD.99.122002}{\JournalTitle{\prd}, 99,
  122002}

\bibitem[{{Arzoumanian} {et~al.}(2002){Arzoumanian}, {Chernoff}, \&
  {Cordes}}]{arzoumanian:2002}
{Arzoumanian}, Z., {Chernoff}, D.~F., \& {Cordes}, J.~M. 2002,
  \href{http://dx.doi.org/10.1086/338805}{\JournalTitle{\apj}, 568, 289}

\bibitem[{Astone {et~al.}(2014)Astone, Colla, D'Antonio, Frasca, \&
  Palomba}]{FH}
Astone, P., Colla, A., D'Antonio, S., Frasca, S., \& Palomba, C. 2014,
  \href{http://dx.doi.org/10.1103/PhysRevD.90.042002}{\JournalTitle{\prd}, 90,
  042002}

\bibitem[{{Bhattacharyya}(2021)}]{Bhattacharyya:2021}
{Bhattacharyya}, S. 2021,
  \href{http://dx.doi.org/10.1093/mnrasl/slab001}{\JournalTitle{\mnras}, 502,
  L45}

\bibitem[{{Bildsten}(1998)}]{bildsten:1998}
{Bildsten}, L. 1998,
  \href{http://dx.doi.org/10.1086/311440}{\JournalTitle{\apjl}, 501, L89}

\bibitem[{{Binney} \& {Merrifield}(1998)}]{Binney:1998}
{Binney}, J., \& {Merrifield}, M. 1998, {Galactic Astronomy}

\bibitem[{{Binney} \& {Tremaine}(2008)}]{Dynamics}
{Binney}, J., \& {Tremaine}, S. 2008, {Galactic Dynamics: Second Edition}

\bibitem[{Caride {et~al.}(2019)Caride, Inta, Owen, \&
  Rajbhandari}]{PhysRevD.100.064013}
Caride, S., Inta, R., Owen, B.~J., \& Rajbhandari, B. 2019,
  \href{http://dx.doi.org/10.1103/PhysRevD.100.064013}{\JournalTitle{\prd},
  100, 064013}

\bibitem[{Covas \& Sintes(2019)}]{Covas:2019}
Covas, P.~B., \& Sintes, A.~M. 2019,
  \href{http://dx.doi.org/10.1103/PhysRevD.99.124019}{\JournalTitle{\prd}, 99,
  124019}

\bibitem[{{Dergachev} \& {Alessandra Papa}(2021)}]{Dergachev_Papa:2021b}
{Dergachev}, V., \& {Alessandra Papa}, M. 2021, \JournalTitle{arXiv e-prints},
  arXiv:2104.09007

\bibitem[{{Dergachev} \& {Papa}(2020)}]{dergachev:2020}
{Dergachev}, V., \& {Papa}, M.~A. 2020,
  \href{http://dx.doi.org/10.1103/PhysRevLett.125.171101}{\JournalTitle{\prl},
  125, 171101}

\bibitem[{{Dergachev} \& {Papa}(2021)}]{Dergachev_Papa:2021a}
---. 2021,
  \href{http://dx.doi.org/10.1103/PhysRevD.103.063019}{\JournalTitle{\prd},
  103, 063019}

\bibitem[{{Diehl} {et~al.}(2006){Diehl}, {Halloin}, {Kretschmer}, {Lichti},
  {Sch{\"o}nfelder}, {Strong}, {von Kienlin}, {Wang}, {Jean}, {Kn{\"o}dlseder},
  {Roques}, {Weidenspointner}, {Schanne}, {Hartmann}, {Winkler}, \&
  {Wunderer}}]{diehl:2006}
{Diehl}, R., {Halloin}, H., {Kretschmer}, K., {et~al.} 2006,
  \href{http://dx.doi.org/10.1038/nature04364}{\JournalTitle{\nat}, 439, 45}

\bibitem[{{Dwyer} {et~al.}(2015){Dwyer}, {Sigg}, {Ballmer}, {Barsotti},
  {Mavalvala}, \& {Evans}}]{dwyer:2015}
{Dwyer}, S., {Sigg}, D., {Ballmer}, S.~W., {et~al.} 2015,
  \href{http://dx.doi.org/10.1103/PhysRevD.91.082001}{\JournalTitle{\prd}, 91,
  082001}

\bibitem[{{Faucher-Gigu{\`e}re} \& {Loeb}(2010)}]{faucher:2010}
{Faucher-Gigu{\`e}re}, C.-A., \& {Loeb}, A. 2010,
  \href{http://dx.doi.org/10.1088/1475-7516/2010/01/005}{\JournalTitle{\jcap},
  2010, 005}

\bibitem[{Gittins {et~al.}(2020)Gittins, Andersson, \&
  Jones}]{10.1093/mnras/staa3635}
Gittins, F., Andersson, N., \& Jones, D.~I. 2020,
  \href{http://dx.doi.org/10.1093/mnras/staa3635}{\JournalTitle{Monthly Notices
  of the Royal Astronomical Society}, 500, 5570}

\bibitem[{{Gravity Collaboration} {et~al.}(2019){Gravity Collaboration},
  {Abuter}, {Amorim}, {Baub{\"o}ck}, {Berger}, {Bonnet}, {Brandner},
  {Cl{\'e}net}, {Coud{\'e} Du Foresto}, {de Zeeuw}, {Dexter}, {Duvert},
  {Eckart}, {Eisenhauer}, {F{\"o}rster Schreiber}, {Garcia}, {Gao}, {Gendron},
  {Genzel}, {Gerhard}, {Gillessen}, {Habibi}, {Haubois}, {Henning}, {Hippler},
  {Horrobin}, {Jim{\'e}nez-Rosales}, {Jocou}, {Kervella}, {Lacour},
  {Lapeyr{\`e}re}, {Le Bouquin}, {L{\'e}na}, {Ott}, {Paumard}, {Perraut},
  {Perrin}, {Pfuhl}, {Rabien}, {Rodriguez Coira}, {Rousset}, {Scheithauer},
  {Sternberg}, {Straub}, {Straubmeier}, {Sturm}, {Tacconi}, {Vincent}, {von
  Fellenberg}, {Waisberg}, {Widmann}, {Wieprecht}, {Wiezorrek}, {Woillez}, \&
  {Yazici}}]{galcenter-dist}
{Gravity Collaboration}, {Abuter}, R., {Amorim}, A., {et~al.} 2019,
  \href{http://dx.doi.org/10.1051/0004-6361/201935656}{\JournalTitle{\aap},
  625, L10}

\bibitem[{{Hessels} {et~al.}(2006){Hessels}, {Ransom}, {Stairs}, {Freire},
  {Kaspi}, \& {Camilo}}]{Hessels:2006}
{Hessels}, J. W.~T., {Ransom}, S.~M., {Stairs}, I.~H., {et~al.} 2006,
  \href{http://dx.doi.org/10.1126/science.1123430}{\JournalTitle{Science}, 311,
  1901}

\bibitem[{{Hobbs} {et~al.}(2020){Hobbs}, {Manchester}, \& {Toomey}}]{ATNF-site}
{Hobbs}, G., {Manchester}, R., \& {Toomey}, L. 2020, The ATNF Pulsar Database

\bibitem[{Horowitz \& Kadau(2009)}]{PhysRevLett.102.191102}
Horowitz, C.~J., \& Kadau, K. 2009,
  \href{http://dx.doi.org/10.1103/PhysRevLett.102.191102}{\JournalTitle{\prl},
  102, 191102}

\bibitem[{Jaranowski {et~al.}(1998)Jaranowski, Kr\'olak, \& Schutz}]{TDFstat}
Jaranowski, P., Kr\'olak, A., \& Schutz, B.~F. 1998,
  \href{http://dx.doi.org/10.1103/PhysRevD.58.063001}{\JournalTitle{\prd}, 58,
  063001}

\bibitem[{{Johnson-McDaniel} \& {Owen}(2013)}]{johnson-mcdaniel:2013}
{Johnson-McDaniel}, N.~K., \& {Owen}, B.~J. 2013,
  \href{http://dx.doi.org/10.1103/PhysRevD.88.044004}{\JournalTitle{\prd}, 88,
  044004}

\bibitem[{{Katsuda} {et~al.}(2018){Katsuda}, {Morii}, {Janka},
  {Wongwathanarat}, {Nakamura}, {Kotake}, {Mori}, {M{\"u}ller}, {Takiwaki},
  {Tanaka}, {Tominaga}, \& {Tsunemi}}]{katsuda:2018}
{Katsuda}, S., {Morii}, M., {Janka}, H.-T., {et~al.} 2018,
  \href{http://dx.doi.org/10.3847/1538-4357/aab092}{\JournalTitle{\apj}, 856,
  18}

\bibitem[{Krishnan {et~al.}(2004)Krishnan, Sintes, Papa, Schutz, Frasca, \&
  Palomba}]{Covas:2004}
Krishnan, B., Sintes, A.~M., Papa, M.~A., {et~al.} 2004,
  \href{http://dx.doi.org/10.1103/PhysRevD.70.082001}{\JournalTitle{\prd}, 70,
  082001}

\bibitem[{{Lasky}(2015)}]{lasky:2015}
{Lasky}, P.~D. 2015,
  \href{http://dx.doi.org/10.1017/pasa.2015.35}{\JournalTitle{\pasa}, 32, e034}

\bibitem[{{Manchester} {et~al.}(2005){Manchester}, {Hobbs}, {Teoh}, \&
  {Hobbs}}]{ATNF}
{Manchester}, R.~N., {Hobbs}, G.~B., {Teoh}, A., \& {Hobbs}, M. 2005,
  \href{http://dx.doi.org/10.1086/428488}{\JournalTitle{\aj}, 129, 1993}

\bibitem[{{Nakamura} {et~al.}(2019){Nakamura}, {Takiwaki}, \&
  {Kotake}}]{nakamura:2019}
{Nakamura}, K., {Takiwaki}, T., \& {Kotake}, K. 2019,
  \href{http://dx.doi.org/10.1093/pasj/psz080}{\JournalTitle{\pasj}, 71, 98}

\bibitem[{{Owen}(2005)}]{owen:2005}
{Owen}, B.~J. 2005,
  \href{http://dx.doi.org/10.1103/PhysRevLett.95.211101}{\JournalTitle{\prl},
  95, 211101}

\bibitem[{{Papitto} {et~al.}(2013){Papitto}, {Hessels}, {Burgay}, {Ransom},
  {Rea}, {Possenti}, {Stairs}, {Ferrigno}, \& {Bozz}}]{Papitto:2013}
{Papitto}, A., {Hessels}, J.~W.~T., {Burgay}, M., {et~al.} 2013,
  \JournalTitle{The Astronomer's Telegram}, 5069, 1

\bibitem[{{Punturo} {et~al.}(2010){Punturo}, {Abernathy}, {Acernese}, {Allen},
  {Andersson}, {Arun}, {Barone}, {Barr}, {Barsuglia}, {Beker}, {Beveridge},
  {Birindelli}, {Bose}, {Bosi}, {Braccini}, {Bradaschia}, {Bulik}, {Calloni},
  {Cella}, {Chassande Mottin}, {Chelkowski}, {Chincarini}, {Clark}, {Coccia},
  {Colacino}, {Colas}, {Cumming}, {Cunningham}, {Cuoco}, {Danilishin},
  {Danzmann}, {De Luca}, {De Salvo}, {Dent}, {De Rosa}, {Di Fiore}, {Di
  Virgilio}, {Doets}, {Fafone}, {Falferi}, {Flaminio}, {Franc}, {Frasconi},
  {Freise}, {Fulda}, {Gair}, {Gemme}, {Gennai}, {Giazotto}, {Glampedakis},
  {Granata}, {Grote}, {Guidi}, {Hammond}, {Hannam}, {Harms}, {Heinert},
  {Hendry}, {Heng}, {Hennes}, {Hild}, {Hough}, {Husa}, {Huttner}, {Jones},
  {Khalili}, {Kokeyama}, {Kokkotas}, {Krishnan}, {Lorenzini}, {L{\"u}ck},
  {Majorana}, {Mandel}, {Mandic}, {Martin}, {Michel}, {Minenkov}, {Morgado},
  {Mosca}, {Mours}, {M{\"u}ller─Ebhardt}, {Murray}, {Nawrodt}, {Nelson},
  {Oshaughnessy}, {Ott}, {Palomba}, {Paoli}, {Parguez}, {Pasqualetti},
  {Passaquieti}, {Passuello}, {Pinard}, {Poggiani}, {Popolizio}, {Prato},
  {Puppo}, {Rabeling}, {Rapagnani}, {Read}, {Regimbau}, {Rehbein}, {Reid},
  {Rezzolla}, {Ricci}, {Richard}, {Rocchi}, {Rowan}, {R{\"u}diger}, {Sassolas},
  {Sathyaprakash}, {Schnabel}, {Schwarz}, {Seidel}, {Sintes}, {Somiya},
  {Speirits}, {Strain}, {Strigin}, {Sutton}, {Tarabrin}, {Th{\"u}ring}, {van
  den Brand}, {van Leewen}, {van Veggel}, {van den Broeck}, {Vecchio},
  {Veitch}, {Vetrano}, {Vicere}, {Vyatchanin}, {Willke}, {Woan}, {Wolfango}, \&
  {Yamamoto}}]{punturo:2010}
{Punturo}, M., {Abernathy}, M., {Acernese}, F., {et~al.} 2010,
  \href{http://dx.doi.org/10.1088/0264-9381/27/19/194002}{\JournalTitle{Classical
  and Quantum Gravity}, 27, 194002}

\bibitem[{{Radhakrishnan} \& {Srinivasan}(1982)}]{Radhakrishnan:1982}
{Radhakrishnan}, V., \& {Srinivasan}, G. 1982, \JournalTitle{Current Science},
  51, 1096

\bibitem[{{Riles}(2017)}]{riles:2017}
{Riles}, K. 2017,
  \href{http://dx.doi.org/10.1142/S021773231730035X}{\JournalTitle{Modern
  Physics Letters A}, 32, 1730035}

\bibitem[{{Sartore} {et~al.}(2010){Sartore}, {Ripamonti}, {Treves}, \&
  {Turolla}}]{sartore:2010}
{Sartore}, N., {Ripamonti}, E., {Treves}, A., \& {Turolla}, R. 2010,
  \href{http://dx.doi.org/10.1051/0004-6361/200912222}{\JournalTitle{\aap},
  510, A23}

\bibitem[{{SciPy Community}(2021)}]{scipy.kde}
{SciPy Community}. 2021, SciPy v1.6.2 Reference Guide,
  scipy.stats.gaussian\_kde documentation

\bibitem[{{Scott}(2015)}]{scottsrule}
{Scott}, D.~W. 2015, {Multivariate Density Estimation: Theory, Practice, and
  Visualization, 2nd Edition}

\bibitem[{{Shklovskii}(1970)}]{shklovskii:1970}
{Shklovskii}, I.~S. 1970, \JournalTitle{\sovast}, 13, 562

\bibitem[{{Steltner} {et~al.}(2021){Steltner}, {Papa}, {Eggenstein}, {Allen},
  {Dergachev}, {Prix}, {Machenschalk}, {Walsh}, {Zhu}, {Behnke}, \&
  {Kwang}}]{steltner2020einsteinhome}
{Steltner}, B., {Papa}, M.~A., {Eggenstein}, H.~B., {et~al.} 2021,
  \href{http://dx.doi.org/10.3847/1538-4357/abc7c9}{\JournalTitle{\apj}, 909,
  79}

\bibitem[{{Taani} {et~al.}(2012){Taani}, {Naso}, {Wei}, {Zhang}, \&
  {Zhao}}]{Taani:2012}
{Taani}, A., {Naso}, L., {Wei}, Y., {Zhang}, C., \& {Zhao}, Y. 2012,
  \href{http://dx.doi.org/10.1007/s10509-012-1121-7}{\JournalTitle{\apss}, 341,
  601}

\bibitem[{{Ushomirsky} {et~al.}(2000){Ushomirsky}, {Cutler}, \&
  {Bildsten}}]{ushomirsky:2000}
{Ushomirsky}, G., {Cutler}, C., \& {Bildsten}, L. 2000,
  \href{http://dx.doi.org/10.1046/j.1365-8711.2000.03938.x}{\JournalTitle{\mnras},
  319, 902}

\bibitem[{{Virtanen} {et~al.}(2020){Virtanen}, {Gommers}, {Oliphant},
  {Haberland}, {Reddy}, {Cournapeau}, {Burovski}, {Peterson}, {Weckesser},
  {Bright}, {van der Walt}, {Brett}, {Wilson}, {Jarrod Millman}, {Mayorov},
  {Nelson}, {Jones}, {Kern}, {Larson}, {Carey}, {Polat}, {Feng}, {Moore}, {Vand
  erPlas}, {Laxalde}, {Perktold}, {Cimrman}, {Henriksen}, {Quintero}, {Harris},
  {Archibald}, {Ribeiro}, {Pedregosa}, {van Mulbregt}, \&
  {Contributors}}]{scipy}
{Virtanen}, P., {Gommers}, R., {Oliphant}, T.~E., {et~al.} 2020,
  \href{http://dx.doi.org/https://doi.org/10.1038/s41592-019-0686-2}{\JournalTitle{Nature
  Methods}, 17, 261}

\bibitem[{{Wijnands} \& {van der Klis}(1998)}]{Wijnands:1998}
{Wijnands}, R., \& {van der Klis}, M. 1998,
  \href{http://dx.doi.org/10.1038/28557}{\JournalTitle{\nat}, 394, 344}

\bibitem[{{Woan} {et~al.}(2018){Woan}, {Pitkin}, {Haskell}, {Jones}, \&
  {Lasky}}]{woan:2018}
{Woan}, G., {Pitkin}, M.~D., {Haskell}, B., {Jones}, D.~I., \& {Lasky}, P.~D.
  2018, \href{http://dx.doi.org/10.3847/2041-8213/aad86a}{\JournalTitle{\apjl},
  863, L40}

\bibitem[{{Zimmermann} \& {Szedenits}(1979)}]{zimmerman:1979}
{Zimmermann}, M., \& {Szedenits}, E., J. 1979,
  \href{http://dx.doi.org/10.1103/PhysRevD.20.351}{\JournalTitle{\prd}, 20,
  351}

\end{thebibliography}

\end{document}